\documentclass[conference,compsoc]{IEEEtran}

\usepackage[nocompress]{cite}
\usepackage{amsmath,amssymb,amsfonts}
\usepackage{graphicx}
\usepackage[table,x11names]{xcolor}
\usepackage{booktabs}
\usepackage{multirow}
\usepackage{array}
\usepackage{siunitx}
\usepackage{xspace}
\usepackage{pifont}
\usepackage{enumitem}
\usepackage{csquotes}
\usepackage{colortbl}
\usepackage{float}
\usepackage{stfloats} 
\usepackage{algorithm}
\usepackage{algpseudocode}
\usepackage{makecell}
\usepackage{tabularx}
\usepackage[most]{tcolorbox}
\usepackage{listings}
\usepackage[hidelinks]{hyperref}
\usepackage[capitalize]{cleveref}
\usetikzlibrary{positioning}

\tcbuselibrary{breakable,skins,listings}

\newcolumntype{Y}{>{\centering\arraybackslash}X}
\newcommand{\sys}{Matryoshka\xspace}
\newcommand{\vivian}[1]{}
\newcommand{\vp}[1]{}
\newcommand{\jp}[1]{}
\newcommand{\raluca}[1]{}
\newcommand{\eat}[1]{}

\newcommand{\pref}[1]{\S\ref{#1}}

\begin{document}

\title{Semantic-Aware Parsing for Security Logs}

\author{
\IEEEauthorblockN{
Julien Piet\IEEEauthorrefmark{1}\IEEEauthorrefmark{2},
Vivian Fang\IEEEauthorrefmark{1},
Rishi Khare\IEEEauthorrefmark{1},
Scott Coull\IEEEauthorrefmark{2},
Vern Paxson\IEEEauthorrefmark{1}\IEEEauthorrefmark{3},
Raluca Ada Popa\IEEEauthorrefmark{1},
David Wagner\IEEEauthorrefmark{1}}
\IEEEauthorblockA{
\IEEEauthorrefmark{1}University of California, Berkeley \quad
\IEEEauthorrefmark{2}Google \quad
\IEEEauthorrefmark{3}Corelight
}
}

\maketitle

\begin{abstract}
Security logs are foundational to threat detection and post-incident investigation, yet analysts often struggle to fully leverage them due to their heterogeneity and unstructured nature. The standard practice of manually writing parsers to normalize the data in security event management systems is time-consuming and costly due to the long tail of log formats. Meanwhile, querying raw logs without explicit parsing using large language models (LLMs) is impractical at scale.

In this paper, we introduce \sys, an end-to-end system leveraging LLMs to automatically generate semantically-aware structured log parsers without labeled examples or human intervention. \sys achieves this by directly inferring log syntax, variable naming, and normalization to common security-specific schemas (e.g., OCSF~\cite{ocsf}) from unlabeled log line samples, then generating deterministic parsers and mapping rules that can be efficiently applied during data ingest. This approach provides analysts with semantically-rich data representations at scale, facilitating rapid and precise log search without the traditional burden of manual parser construction.

We evaluate \sys's capabilities through both established template generation datasets and new datasets curated to establish end-to-end performance on a realistic distribution of log types. Our experiments show that \sys outperforms prior work on syntax parsing while matching human-generated parsers in both side-by-side comparisons and retrieval for security-relevant queries. These results demonstrate that \sys significantly reduces manual effort by automatically extracting and organizing valuable security data, moving us closer to fully automated, AI-driven analytics.
\end{abstract}

\section{Introduction}
\label{introduction}

Security operations depend on analysts' ability to rapidly search and cross-correlate vast quantities of data to detect and respond to threats. However, this comes with a fundamental challenge: correlating events from multiple log sources (e.g., infrastructure operational events, network devices, applications, and security tools) that are heterogeneous, massive, and most often \emph{semi-structured}. Each data source generates logs in different formats with varying levels of detail, and even within a given source these formats come in many variants and evolve over time. 

Logs underpin two core security operations workflows: (i) interactive investigations that reconstruct a chain of events, and (ii) automated real-time detection using rules and machine learning. To leverage the sources of log data, security teams write parsers that convert logs into a structured schema, such as Google's Unified Data Model~\cite{udm} (UDM) or the Open Cybersecurity Schema Framework~\cite{ocsf} (OCSF). These schemas aim to unify the representation of log events across sources and provide a structured, normalized way to represent this data used in security event management systems, but they are difficult to maintain at scale.

To underscore this challenge, UDM exposes around 20,000 distinct attributes, while OCSF provides more than 50,000. A typical event management platform, such as Google SecOps, provides over 1,000 distinct log parsers~\cite{googleparsers}, and pushes more than 200 parser updates monthly, requiring more than 4,000 engineer-hours. Despite this effort, mappings are often ambiguous or incomplete, leading to poor detection and investigation outcomes. In our evaluation of end-to-end query performance (\S~\ref{ssec:e2e-def}), for instance, human-written parsers achieved a precision of only 0.50 and recall of 0.48 across a diverse set of security-relevant queries due to the inherent difficulty in producing normalized schema mappings.

On the surface, modern large language model~(LLM)-based systems seem ideally suited for this problem. LLMs can understand system logs~\cite{liu2025loglmtaskbasedinstructionbasedautomated, cui2024logevalcomprehensivebenchmarksuite, karlsen2024}, retrieval-augmented generation~(RAG) systems can query heterogeneous data using natural language~\cite{rag}, and they can even generate SQL queries for structured data analysis~\cite{hong2024next, yu2018syntaxsqlnetsyntaxtreenetworks, yu2019spiderlargescalehumanlabeleddataset, guo2019complextexttosqlcrossdomaindatabase, scholak2021picardparsingincrementallyconstrained}. However, real-world systems generate millions of log lines daily---far exceeding any language model's context window. Log data often combines natural language, technical identifiers, and structured delimiters, making standard text embeddings inefficient and inaccurate. Even recent advances in querying unstructured data cannot keep pace with log generation rates since those approaches require embedding each new line individually and often resort to imprecise clustering for speed~\cite{dai2024uqequeryengineunstructured}. 

Security teams are thus left with two suboptimal approaches: either (1)~write parsers by hand to convert them to a structured form, requiring substantial human resources but enabling efficient ingestion and search over logs, or (2)~use AI search tools to retrieve log events from unstructured data formats, a slow and costly approach that can result in poor retrieval and fails to scale to security workload volumes, thereby sacrificing both ingestion speed and quality.

Instead, we propose \sys, the first end-to-end system leveraging LLMs to automatically generate \emph{deterministic, semantically-aware structured log parsers}, offering fast ingestion and search with no human intervention. \sys proceeds in three steps. First, a \emph{syntax parser} captures the syntax of each log line and identifies variables. Second, a \emph{semantic naming} step consistently names and clusters these extracted variables, producing a coherent schema suitable for structured queries. Finally, and optionally, \sys can map the newly created fields to normalized security schemas, such as UDM or OCSF. At run time our generated parsers rely exclusively on static regular-expression matching---\emph{not} LLM-based analysis---enabling efficient ingestion. The schema produced by the syntax parser and semantic naming provides analysts with a queryable, structured dataset that supports fast searches even without normalization, and the final mapping step enables query and rule creation across log sources using normalized schemas within security event management systems.


\sys addresses three key challenges that make automated end-to-end log parsing difficult in practice:
\begin{enumerate}
[leftmargin=*]
    \item \textbf{Scale and heterogeneity.} Real-world logs are massive, diverse, and evolve constantly, making manual parser maintenance costly and brittle. \sys automates this process by learning log syntax and structure and generating deterministic parsers directly from unlabeled log samples, scaling across heterogeneous formats without human input.
    \item \textbf{Limited expressiveness.} Past work focuses on \emph{template generation}, a different task from log parsing. These methods construct templates with wildcards to match variables, and identify specific log formats~\cite{jiang2024lilac, jiang2024large, brain23, Drain}. Such templates blur boundaries between adjacent fields and often over-capture, leading to poor variable extraction and degraded end-to-end parsing performance. In contrast, \sys's \emph{syntax parsing} unifies log templates into a single parsing tree, relying on regular expressions to accurately capture variable boundaries, and thus enabling the precision necessary for later parsing stages and, ultimately, downstream analysis tasks.
    \item \textbf{Missing semantics.} Prior approaches do not attach semantic names to variables and thus provide no schema for querying. Naively employing a language model for naming yields inconsistent field names (e.g., one log line may use ``\texttt{source\_ip}'' while another uses ``\texttt{src\_ip}''). \sys's \emph{semantic naming} and \emph{mapping} stages incorporate an execution-guided validation phase that enforces global consistency across templates, yielding coherent schema and substantially improving query accuracy.
\end{enumerate}

To evaluate \sys, we use three datasets: (i) SecurityLogs, a new dataset containing five common log types to evaluate schema mapping and end-to-end retrieval, (ii) LogHub2.0~\cite{loghub}, a standard template generation benchmark to demonstrate syntax parsing capabilities, and (iii) examples from 27 real-world enterprise log types from Google SecOps that capture the long tail of log formats seen in practice. Our experiments highlight the efficiency and performance of our approach at each stage in the parsing process. 
On end-to-end retrieval tasks for representative security queries, for instance, \sys demonstrates significant improvements over human-generated parsers in both precision (0.60 vs. 0.50) and recall (0.60 vs. 0.48) on data normalized to the UDM schema. In side-by-side comparisons, \sys's mapping to UDM achieves the same quality as the human-written parsers. Meanwhile, without mapping to a normalized schema (i.e., leveraging semantic naming alone), \sys achieves near-perfect performance on those same security queries, highlighting the inherent difficulty of mapping to normalized security taxonomies for both humans and machines. More importantly, though, these results demonstrate \sys's ability to achieve human-levels of parser quality without any human intervention, thereby enabling more complete threat detection and incident reconstruction. 


We proceed by examining preliminaries of log parsing in~\cref{background}, followed by related work in~\cref{relwork}. \cref{system} provides a detailed description of \sys's architecture and \cref{eval} presents our evaluation methodology and results demonstrating the efficacy of \sys. Finally,~\cref{discussion} addresses the strengths and limitations of our approach. Our source code and benchmark are publicly available. \footnote{https://github.com/julien-piet/matryoshka}
%

\begin{figure*}[t]
    \centering
    \includegraphics[width=\textwidth]{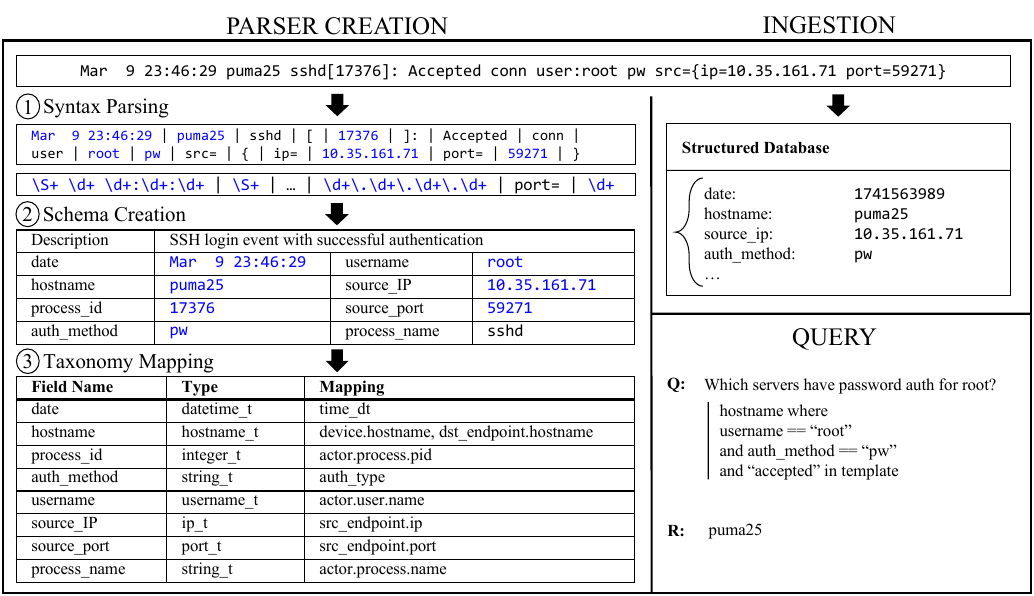}
    \caption{Parsers convert unstructured log lines (top) to structured data suitable for ingestion into a database (upper-right), in a sequence of three steps. The output format makes it easy to query logs for events that satisfy certain conditions (lower-right). \sys creates such parsers by learning the log's syntax~\ding{172}, crafting a schema~\ding{173}, and mapping to a taxonomy~\ding{174}.}
    \label{fig:design-figure}
\end{figure*}

\section{Background}
\label{background}

\subsection{Anatomy of a parser}
System logs are text-based messages that record a program's state and report events, from kernel messages to application activity. Log formats vary widely, even across versions of the same application, complicating parsing and querying. While there is no standard design for log parsers, it is useful to distinguish three conceptual stages that any parser may incorporate: syntax parsing, semantic naming, and schema mapping.

\smallskip\noindent\textbf{Syntax parsing.} Programs emit log lines by interpolating variable entities into a fixed message template. The syntax of each line is thus determined by its template. Templates are defined as sequences of tokens, where each token is either a fixed string or a variable. We associate each variable with a regular expression. Step~\ding{192} in~\cref{fig:design-figure} shows an example template that captures an SSH accepted connection event. 


Many templates may share a common prefix. For instance, Linux system logs often share a prefix indicating the date, hostname, process name, and process ID, followed by program-specific content. Syntax parsing in \sys relies on a tree of templates, where all templates in a subtree share a common prefix. This design helps parse log lines consistently, as repeated prefixes are parsed identically.


For example, consider the log line in~\cref{fig:design-figure}, which highlights some of the challenges of real-world log parsing. Delimiters (\texttt{=}, \texttt{:}, \texttt{\{\}}) mixed with free text and nested key-value pairs make fields hard to separate; for example, the variable ``\texttt{pw}'' implies password authentication without stating it explicitly. Moreover, different clients (or even versions of the same client) may change the syntax or semantics of these log lines in subtle ways.

One possible template for the log line is ``\texttt{\small <*> <*> sshd[<*>]: Accepted conn user:<*> <*> src=\{ip=<*> port=<*>\}}'', and it can be captured by the main branch in the parsing tree represented in~\cref{fig:parsing-tree}. The start of the line (up to the process ID brackets) is present in all log lines, so it represents a branching point in the tree of templates.
Each leaf represents a distinct template.
Because the date and hostname are adjacent, parsing with wildcards would be unable to disambiguate where the date ends and hostname begins.
This motivates the use of regular expressions to capture variables (e.g., ``\texttt{\textbackslash S+\textbackslash s+\textbackslash d+\textbackslash s+\textbackslash d+:\textbackslash d+:\textbackslash d+}'' for the date) and enable unique disambiguation.

\smallskip\noindent\textbf{Semantic naming.} While templates capture a line's syntax, they provide no information about its meaning. Therefore, the next parsing stage maps each template to a schema. Each schema includes a description of the template and a set of fields representing the template's variables and constants of importance, where each field is assigned a name and description. Names must reflect not only the data type of the original variable but also the variable's purpose within the broader log context (see \cref{fig:design-figure}, step~\ding{193}).
After this stage, each log line is mapped to a JSON object containing a description of the template and a list of named fields corresponding to the variables in the line.
Schema for related templates must be consistent (e.g., use the same field name when the ``same'' variable appears in multiple templates).

\smallskip\noindent\textbf{Schema mapping.} The first two stages produce a structured representation of the log file, enabling fast and efficient querying. The third step maps this structured representation to an existing normalized taxonomy so analysts can search over standardized attribute names (see \cref{fig:design-figure}, step \ding{194}). Each named entity in the parser's parsing tree is assigned to one or more attributes in the target taxonomy. Some fields cannot be mapped to standardized attributes. Normalized taxonomies are often incomplete and cannot represent every domain-specific value. In these cases, the field is described using the custom schema from step \ding{193}.

\begin{figure*}[t]
\centering
\begin{tikzpicture}[
font=\ttfamily\scriptsize,        
node distance=0.15cm,             
>=stealth,
normalNode/.style={
draw=black,
rectangle,
rounded corners,
minimum height=1em,
inner sep=1.5pt,
text=black,
align=center
},
varNode/.style={
draw=blue,
rectangle,
rounded corners,
minimum height=1em,
inner sep=1.5pt,
text=blue,
align=center
},
addedNode/.style={
dotted
},
solidEdge/.style={
->,
thick,
draw=black
},
dottedEdge/.style={
->,
thick,
dotted,
draw=black
}
]

\node[varNode] (mar) {Mar  9 23:46:29 \\ \textcolor{blue!60}{\tiny \textbackslash S+\textbackslash s+ $\cdots$ \textbackslash d+:\textbackslash d+}};
\node[varNode,right=0.15cm of mar] (puma) {puma25 \\ \textcolor{blue!60}{\tiny \textbackslash S+}};
\node[normalNode,right=0.15cm of puma] (sshd) {sshd \\ \phantom{\tiny x}};
\node[normalNode,right=0.15cm of sshd] (lb) {[ \\ \phantom{\tiny x}};
\node[varNode,right=0.15cm of lb] (pid) {17376 \\ \textcolor{blue!60}{\tiny \textbackslash d+}};
\node[normalNode,right=0.15cm of pid] (rb) {]: \\ \phantom{\tiny x}};
\node[normalNode,right=0.15cm of rb] (acc) {Accepted \\ \phantom{\tiny x}};
\node[normalNode,right=0.15cm of acc] (conn) {conn \\ \phantom{\tiny x}};
\node[normalNode,right=0.15cm of conn] (userlbl) {user: \\ \phantom{\tiny x}};
\node[varNode,right=0.15cm of userlbl] (root1) {root \\ \textcolor{blue!60}{\tiny \textbackslash S+}};
\node[varNode,right=0.15cm of root1] (pw) {pw \\ \textcolor{blue!60}{\tiny \textbackslash S+}};
\node[normalNode,right=0.15cm of pw] (src) {src= \\ \phantom{\tiny x}};
\node[normalNode,right=0.15cm of src] (lbrace) {\{ \\ \phantom{\tiny x}};
\node[normalNode,right=0.15cm of lbrace] (ipkey) {ip= \\ \phantom{\tiny x}};
\node[varNode,right=0.15cm of ipkey] (ip) {10.35.161.71 \\ \textcolor{blue!60}{\tiny \textbackslash d+.\textbackslash d+.\textbackslash d+.\textbackslash d+}};
\node[normalNode,right=0.15cm of ip] (portkey) {port= \\ \phantom{\tiny x}};
\node[varNode,right=0.15cm of portkey] (prtN) {59271 \\ \textcolor{blue!60}{\tiny \textbackslash d+}};
\node[normalNode,right=0.15cm of prtN] (rbrace) {\} \\ \phantom{\tiny x}};

\node[normalNode,addedNode,above right=0.6cm and 0.15cm of rb] (fail) {Failed \\ \phantom{\tiny x}};
\node[varNode,addedNode,right=0.15cm of fail] (none) {none \\ \textcolor{blue!60}{\tiny \textbackslash S+}};
\node[normalNode,addedNode,right=0.15cm of none] (for2) {for \\ \phantom{\tiny x}};
\node[normalNode,addedNode,right=0.15cm of for2] (usr) {user \\ \phantom{\tiny x}};
\node[varNode,addedNode,right=0.15cm of usr] (root2) {root \\ \textcolor{blue!60}{\tiny \textbackslash S+}};

\draw[solidEdge] (mar) -- (puma);
\draw[solidEdge] (puma) -- (sshd);
\draw[solidEdge] (sshd) -- (lb);
\draw[solidEdge] (lb)   -- (pid);
\draw[solidEdge] (pid)  -- (rb);
\draw[solidEdge] (rb)   -- (acc);
\draw[solidEdge] (acc)  -- (conn);
\draw[solidEdge] (conn) -- (userlbl);
\draw[solidEdge] (userlbl) -- (root1);
\draw[solidEdge] (root1) -- (pw);
\draw[solidEdge] (pw)   -- (src);
\draw[solidEdge] (src)  -- (lbrace);
\draw[solidEdge] (lbrace) -- (ipkey);
\draw[solidEdge] (ipkey) -- (ip);
\draw[solidEdge] (ip)   -- (portkey);
\draw[solidEdge] (portkey) -- (prtN);
\draw[solidEdge] (prtN) -- (rbrace);

\draw[dottedEdge] (rb) -- (fail);
\draw[dottedEdge] (fail) -- (none);
\draw[dottedEdge] (none) -- (for2);
\draw[dottedEdge] (for2) -- (usr);
\draw[dottedEdge] (usr) -- (root2);

\end{tikzpicture}
\caption{Example parsing tree, with two templates that share a common prefix. Each node/token represents either a string constant (black) or a variable (with associated regular expression; blue), and each leaf corresponds to a template (given by the path from the root to that leaf).}
\label{fig:parsing-tree}
\end{figure*}

\subsection{Security operations}
\label{ssec:security_operations}

Security operations teams focus on detecting, investigating, and mitigating threats. Logs underpin two core workflows: \emph{(i) interactive investigations} to reconstruct chains of events, and \emph{(ii) real-time detection} using alerting rules. Both are hindered when key semantics---``who did what to whom, where, and how''---are hidden in unstructured or semi-structured text.

Analysts often start with substring search (e.g., searching for ``\texttt{Accepted conn}'' in \textit{sshd} logs). This is brittle: different software versions use different wording, and substring search struggles with structured fields like dates. To improve accuracy, analysts can write parsers to convert log data to a structured format, and then search this structured data. In practice, parsers can be extremely complex. For instance, one SSH parser we examined used a 3616-character regex inside a 160-line Python script, focusing solely on extracting the usernames, IPs and timestamps of successful connections. The slow, manual process of writing parsers for each data source and type of log event can consume days or even weeks, taking up analyst time that could instead be dedicated to investigating threats.

Industry mitigates this by \emph{normalizing} logs into large taxonomies such as UDM ($\sim$20K attributes) and OCSF ($>$50K). While such breadth captures important nuance, it also makes mapping onerous and fragile. Google SecOps alone maintains roughly 1,000 active parsers~\cite{googleparsers}, while Splunk lists support for logs generated by 100+ companies with monthly updates~\cite{splunkparsers} and Elastic shows a similar maintenance cadence. Maintaining these parsers is expensive: Google SecOps releases over 200 parser updates each month, consuming about 4,000 engineer-hours from over 30 domain specialists. This level of effort indicates that supporting the long tail of real-world log types can be punishing, with many parsers used by fewer than ten customers.

Even with expert effort, semantic mismatches remain. Taxonomies provide an overwhelming level of nuance, making it difficult to determine which field should be used. For instance, UDM exposes ``\texttt{principal.network.application\_protocol}'', ``\texttt{target.network.application\_protocol}'', and ``\texttt{network.application\_protocol}'', any of which could be used to store the value \enquote{HTTP} in an Apache log. In other cases, mappings depend on perspective. In a \emph{reverse SSH tunnel}, host \texttt{A} initiates an outbound connection to host \texttt{B}, after which \texttt{B} can reach services on \texttt{A}. In a single event, \texttt{A} is simultaneously the \emph{source} (packets originate there), the \emph{initiator} (it opened the session), and once the tunnel is active, the \emph{target} of inbound traffic. Choosing one attribute discards context; populating several creates ambiguity. At the other extreme, even rich taxonomies can \emph{under-capture} domain-specific detail. Both issues hinder detection and retrieval accuracy: across a corpus of queries spanning multiple task categories in the NICCS NICE framework~\cite{niccs_nice_framework}, expert parsers achieve an F1 score of only 0.49 because ambiguous or missing mappings propagate into query logic.

In this paper we propose an end-to-end pipeline that removes the manual burden of writing parsers, achieves expert-level mapping, and additionally supports direct querying through both a custom schema and standardized taxonomies (UDM/OCSF).

\subsection{Parser requirements}
Given the sensitive security use cases that we focus on and the end-to-end nature of the parsers developed by \sys, we have designed our system around four core principles:

\begin{itemize}[leftmargin=*]
\item \textbf{Accuracy.} The structured representation of the logs must be correct: each template should match only a single event. Variable tokens should capture parts of the line that can vary and convey some meaningful information about the event. Template schemas must accurately represent the content of log lines, with meaningful variable names. Fields must only be mapped to standardized attributes that capture their intended role. Lastly, queries executed on the structured data should yield the same result as painstakingly searching the raw logs.
\item \textbf{Completeness.} The structured representation must capture all information in the original log file. Any query that could be executed on the raw logs should be equally feasible on the structured data.
\item \textbf{Consistency.} Variables fulfilling the same role across multiple log lines should be handled identically: they must have the same field names, descriptions, OCSF mappings, and data types. Similarly, templates that are syntactically different but serve the same purpose should have the same schema.
\item \textbf{Run-time efficiency.} The parser we produce should operate efficiently: we should not need to invoke a language model on every log line and queries against structured data should run more efficiently than a linear scan over the raw logs.
\end{itemize}

\noindent
While these five principles guide our system, it is impossible to perfectly meet them all in practice. For instance, perfect accuracy cannot be achieved without complete knowledge of the developer's original intention, and consistency involves a subjective notion of semantic similarity across fields.

\section{Related work}
\label{relwork}

Most of the literature refers to \emph{log parsing} as the task of generating templates that identify variables using wildcard placeholders. The intuition is that applications typically generate log messages with a printf-style (format string) API, and each template should correspond to a unique format string in the source code.
We call this \emph{template generation}. These wildcard templates are useful for identifying log events, but insufficient for ingesting the log message and extracting its variables, as we show in~\cref{ssec:gap-to-prior-work}. Template generation has been studied for decades, and existing approaches can be divided into statistics-based and LLM-based approaches.

\smallskip\noindent\textbf{Statistics-based template generation.}
Earlier work on template generation applies statistical methods to find variables in log messages: they identified variables based on word length, frequency, and other statistical features.
Frequency-based methods~\cite{vaarandi2003data,nagappan2010abstracting,vaarandi2015logcluster,dai2020logram} rely on occurrence frequencies or $n$-gram counts to build templates. Clustering-based techniques~\cite{fu2009execution,hamooni2016logmine,tang2011logsig,shima2016length,mizutani2013incremental} group similar log lines to produce templates. Heuristic-based approaches~\cite{tak2021lognroll,jiang2008abstracting,fu2009execution,Drain,SPINE,Spell,IPLoM,POP,brain23} employ various heuristics or rule-based methods to detect the variable parts of each line. These approaches fail to incorporate semantic information about logs, leading to worse performance than more recent works.

\smallskip\noindent\textbf{LLM-based template generation.}
Recent work has trended towards using LLM-based analysis of logs to improve the quality of generated templates.
Earlier iterations of these approaches cast template generation as a token classification task and trained neural networks~\cite{huo2023semparser,li2023did,liu2022uniparser} to mine structural relationships among tokens from log messages.
More recently, researchers have shown that LLMs are effective at template generation~\cite{le2023log,ma2024librelog,zhang2024lemur,yu2024loggenius,astekin2024comparative, vaarandi2025usinglargelanguagemodels}, especially when leveraging in-context learning~\cite{brown2020language}. DivLog~\cite{xu2024divlog}, for instance, selects examples for developers to label; then when generating a log template for a target log line, it includes the most similar labeled examples in the LLM prompt. LILAC~\cite{jiang2024lilac} builds on DivLog by improving the sampling method when choosing examples for a target log line, and adds a parsing cache to reduce the number of LLM queries. Other approaches focus on reducing the number of LLM queries~\cite{zhong2024logparser,huang2024lunar,xiao2024free} or fine-tune smaller LLMs for better performance~\cite{ma2024llmparser,ma2024luk}.


This past work has focused primarily on generating wildcard templates that identify the variables associated with each template. Notably, these templates do not map the variables into a common schema, and their wildcard representation tend to over-capture and conflate consecutive variables.
By contrast, \sys both extracts syntactic information from logs, but also maps each line to a structured, semantically rich format without labeled examples. 
Moreover, our approach improves generates a regular expression for each variable, thereby improving the quality and precision of the templates in the syntax parsing phase.
Finally, past work has typically not been evaluated on security-focused workloads, because of lack of suitable datasets; we address this by collecting a dataset of security-relevant logs, and discover that real-world security logs are more diverse and challenging than prior datasets.

\smallskip\noindent\textbf{Schema matching.}
The database research community has previously studied schema mapping, where the goal is to map data in one schema into a second schema ~\cite{rahm2001survey,smat}.
Existing methods are effective in settings where both schemas are thoroughly documented, but they are insufficient in our setting where schemas are auto-generated, do not come with documentation of the meaning of fields in the schema, and can contain tens of thousands of fields.
Recent work explores using LLMs for schema mapping~\cite{sheetrit2024rematchretrievalenhancedschema,parciak2024schemamatchinglargelanguage,zhang2024smutfschemamatchingusing, liu2024magnetocombiningsmalllarge, xu2025kcmfknowledgecompliantframeworkschema}, but it too shares some of these limitations.


\smallskip\noindent\textbf{Log querying.}
LLMs are good at analyzing data~\cite{chen2023largelanguagemodelsfew1shot, fang2024largelanguagemodelsllmstabular, li2023tablegpttabletunedgptdiverse}, including unstructured logs~\cite{liu2025loglmtaskbasedinstructionbasedautomated, cui2024logevalcomprehensivebenchmarksuite, karlsen2024}. Structured data can be queried using text-to-SQL tools~\cite{hong2024next, yu2018syntaxsqlnetsyntaxtreenetworks, yu2019spiderlargescalehumanlabeleddataset, guo2019complextexttosqlcrossdomaindatabase, scholak2021picardparsingincrementallyconstrained}, while unstructured data is usually queried using RAG~\cite{gao2024retrievalaugmentedgenerationlargelanguage, rag}. 
LLMs can be used to plan out complex queries over structured data~\cite{liu2025optimizingllmqueriesrelational}, unstructured data~\cite{anderson2024designllmpoweredunstructuredanalytics}, or both ~\cite{dai2024uqequeryengineunstructured}.
Unfortunately, these methods are not sufficient for our use case: they typically apply a LLM to each log line, which is too expensive; query expressivity is limited, because they rely on a vector embedding of each line; and they produce approximate results.


\smallskip\noindent\textbf{Security logs.}
In the security domain, structured logs are usually ingested into log managers such as Splunk~\cite{splunk}, Google Security Operations~\cite{gso}, or Elastic Security~\cite{elastic}. These systems operate over normalized/structured data and support expressive query languages (often with AI-assisted query builders). However, the effectiveness of such queries relies on accurate normalization. Similar events may be recorded in multiple formats and mapped differently across sources, making it challenging to write a query that matches all instances of an event.

\begin{figure*}[t]
    \centering
    \includegraphics[width=\linewidth]{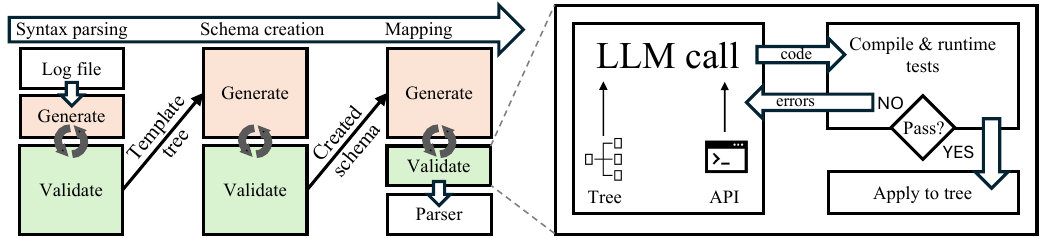}
    \caption{\sys creates parsers in three stages, each subdivided into two logical steps. \sys generates candidate solutions for log lines sequentially, then it validates batches of lines to enforce consistency. Validation produces code to edit the parse tree, iterating until valid.}
    \label{fig:design-figure-2}
\end{figure*}

\section{System architecture}
\label{system}

\sys automates the creation of log parsers without requiring any labeled data or human intervention.\footnote{Named after Matryoshka dolls because of the nested layers of parsing involved in extracting vital information from log messages.} 
Its design mirrors the architecture of a parser: it creates a parsing tree, builds schemas, and maps fields to a taxonomy. 
To achieve this, \sys uses large language models only during parser creation, never at run time. 
This ensures that parsers applied to live logs are fast and deterministic.
As illustrated in~\cref{fig:design-figure-2}, parser construction unfolds in three stages, each following the same parallel two-phase pattern:

\begin{enumerate}[leftmargin=*]
  \item \textbf{Generation.}  Process inputs sequentially, using
        standard LLM techniques (few-shot prompting, guided chain of
        thought, self-consistency) to propose solutions.
  \item \textbf{Validation.}  Revisit candidates in batches,
        reconcile them for global consistency, and fix or discard any
        that fail structural checks.
\end{enumerate}

\noindent
Concretely, this means during \emph{syntax parsing} we first let the LLM generate new templates for unmatched lines, then validate entire sub-trees for consistency; in \emph{schema creation} we name variables line-by-line, then run a batch pass that merges similar concepts into a single canonical field; and in \emph{mapping} we assign provisional attributes one field at a time, then batch-correct them so that twin and sibling fields share consistent parents in the hierarchy. 
This generate-validate loop produces deterministic, consistent parsers and bridges the gap between probabilistic language models and the deterministic behavior required for parsing.
We quantitatively evaluate the contribution of validation in~\cref{ssec:ablations}.

\subsection{Generation phase}
For each new object---an unmatched log line, a template, or a field awaiting a mapping---we run a multi-turn prompt that relies on the following building blocks.

\begin{itemize}[leftmargin=*]
    \item \textbf{Chain-of-thought.} Prompts the model to explain its reasoning, ensuring the model analyzes all relevant context. We supply a structured algorithm to steer its reasoning process, ensuring thorough analysis of the input data.
    \item \textbf{Self-consistency.} We sample several completions using slightly different variants of the prompt and keep the majority vote. Self-consistency has been shown to outperform greedy decoding by allowing the model to explore multiple reasoning paths~\cite{wang2023selfconsistencyimproveschainthought}.
    \item\textbf{Execution checks.} Each generation loop produces a candidate, runs deterministic checks (regex compilation, field existence$\ldots$), and, if failed, provides the model with feedback. This is repeated until all checks pass.
    \item \textbf{Few-shot examples with description embeddings}. Including few-shot examples that closely match the current prompt substantially improves both consistency and accuracy. Crucially, these examples are drawn from previously generated objects, rather than human-provided labels. Standard retrieval approaches perform poorly due to the mix of structured and unstructured text, so we first have the LLM write a short semantic \emph{description} of the example, then embed that description; cosine similarity over these summaries reliably surfaces the most relevant prior examples. By default, we include up to 5 few-shot examples, but performance is similar when using 3 to 7 examples.
\end{itemize}

We now detail the generation process for each of the three parsing stages. This entire process is automatic, with no human involvement. The full set of prompts can be found in our source code.

\subsubsection{Syntax parsing}
\label{ssec:template-generation}
We learn a parsing tree iteratively from raw logs. Unmatched lines are clustered before 
generating templates. Specifically, we buffer up to 2,500 lines and partition 
them in two lightweight passes: (i) a coarse pass that groups lines by their 
longest shared prefix with the current tree, and (ii) a fine pass that clusters 
lines via cosine similarity over embeddings of short, LLM-written semantic descriptions. 
We rely on DBSCAN~\cite{dbscan} ($\epsilon=0.05$). 
The most dense cluster is then fed to a language model to confirm its elements should 
all be parsed by a single template---if not, the model isolates a subset of lines 
from the cluster that share a template.

The language model then proposes a template for this cluster, by providing a 
tokenization into constants and variables and a regex per variable. Proposals
undergo a self-correction loop until each variable's regex compiles, then are 
sampled with self-consistency and scored by a simple objective: maximize 
coverage of the cluster while minimizing spillover onto already-parsed lines. 

Once the model produced a candidate template, we check if it overlaps existing ones. 
Template overlaps can be legitimate, for instance, when an older template is too specific 
or a new one is needed for malformed values. They may also indicate over-capture, in which 
case the new template should be discarded. If a new template overlaps existing ones, we 
sample overlapping lines and ask the model if they come from the same format. If so, the 
old templates are likely too specific; we replace them with the candidate. If not, the new 
template is likely over-capturing; we run a self-correction loop to narrow the candidate. 

The final template is added to the parsing tree; its prompt and language model 
output are cached to be used as few-shot examples in subsequent generations.

\subsubsection{Schema creation}
\label{ssec:schema-creation}
Schemas are learned iteratively from templates. Given a template, the model names 
each variable and semantically relevant constant, then writes a description for each named token. 
Since templates reside in the same parsing tree, their tokens often share common prefixes. We give 
the model the names and descriptions for these existing prefix tokens, and enforce their reuse through self-correction. 

We ensure consistency through few-shot prompting, encouraging the model to reuse variable names and 
descriptions across templates when the same semantic roles are present. We identify similar templates 
that have already been mapped to a schema using description embeddings, and provide them as few-shot examples.

\subsubsection{Mapping}
\label{ssec:schema-mapping}
We flatten the target taxonomy to leaf attributes, use the language model to write a short description of each, 
and embed that description.\footnote{This pre-processing step only needs to be run once for every update to the taxonomy. 
Parser update cycles typically occur every few months.}
We then map fields in our created schema iteratively. For each field, we embed its description and filter the closest 50 
target taxonomy attributes based on cosine similarity of their embeddings. Then, we have the language model select the most appropriate attributes, if any, given these filtered candidates. 

We bias for consistency by injecting sibling fields of already assigned attributes into the filtered list of candidates, and providing previously mapped similar fields as few-shot examples. Sibling fields are attributes that share a direct parent in the taxonomy's hierarchy. For instance, if a field called ``{\tt src\_ip}'' has already been assigned to ``{\tt src\_endpoint.ip}'', we will add sibling fields such as ``{\tt src\_endpoint.hostname}'' to the candidate list of other fields that appear in a template with ``{\tt src\_ip}''. 

Some taxonomies have stronger type systems than others. OCSF, for instance, has types for usernames, dates, and process names. When mapping to such a taxonomy, we optionally have the language model assign each field a type first, then use the types to prune potential candidates in the mapping stage. 

\subsection{Validation phase}
Batch validation lets the model view related candidates side-by-side,
catching inconsistencies that a sequential pass would miss.
As an example, consider the SSH log line in~\cref{fig:design-figure} whose 
authentication method is ``\texttt{pw}'' for password. If the model sees only that line during generation it may incorrectly treat \texttt{pw} as a constant token; when a future line introduces \texttt{pka} and \texttt{mfa}, these will generate another template. Validation merges all three into a single \texttt{auth\_method} variable. 

Because a full parse tree or schema often exceeds context limits, we
validate in \emph{rolling batches}: each batch sees a subset of previously validated
solutions as well as new candidates and may only modify the latter. 
This limits the scope of validation to only change the most recently 
generated solutions, preventing it from fixing some long-distance inconsistencies 
in the input.

Validation requires changing the parsing tree, either to rearrange nodes or add field names. 
Trees are complex objects for language models to work with, and simply asking the model to return a corrected tree almost always leads to mistakes. 
Instead, we frame validation as a code generation task: we supply
API stubs that manipulate the tree (add, delete, move nodes; rename
fields; adjust mappings) and ask the model to output Python code plus a
natural-language rationale.  The code is executed inside a controlled sandbox;
Compile and runtime errors or failed checks an automatic self-repair loop. 

This technique, illustrated in~\cref{fig:design-figure-2}, has three main advantages: (1) it frames validation as a coding problem, which is in-distribution for many language models; (2) correctness of the validated tree can be checked after every API call, errors can be used to give fine-grained feedback to the model so it can refine its code until it produces a valid output; 
(3) we can restrict the API to only allow some actions and prevent the model from changing ground truth parts of the tree. 
This strategy builds on recent ideas in LLM self-verification and code-execution feedback loops~\cite{wang2023selfconsistencyimproveschainthought,chen2023teachinglargelanguagemodels, schick2023toolformerlanguagemodelsteach}, but we adapt them for consistency rather than reasoning accuracy, using controlled code synthesis to repair parser trees deterministically.

In syntax parsing, we let the model add, delete and move nodes in the tree. 
When executing the code, we check that the graph resulting from the change is still a tree and that the final tree still matches the same set of lines as previously. 
In schema creation, we let the model create new field names, assign field names to nodes in the tree, and assign descriptions to field names. We verify new field names are unique. Separating field name creation and assignment avoids the model accidentally naming two different concepts the same name. 
In mapping, we let the model assign field names to attributes, making sure the attribute exists in the target taxonomy. We enforce that at most $N$ attributes are assigned per field (where $N$ is a parameter that defaults to 1). 

This alternating generate/validate strategy enforces consistency by construction. The generation phase relies on small, focused LLM calls that iteratively propose high-quality candidates, while the validation phase executes model-generated repair code inside a sandbox to verify and enforce structural invariants across related log events. Although generation can run on lightweight non-reasoning models, advanced reasoning capabilities are required for validation. This back-and-forth between generation and validation yields deterministic and consistent parsers, bridging the gap between probabilistic language models and the reliability requirements of security infrastructure.

\section{Evaluation}
\label{eval}

To evaluate \sys, we use three different datasets to show our approach produces high quality parsers, comparable to those produced by expert analysts, on a wide variety of log types.

\begin{enumerate}[leftmargin=*]
    \item \textbf{SecurityLogs (\pref{ssec:bugtracker}).} We curated SecurityLogs by crawling system logs uploaded to Red Hat Bugzilla and extracting 500K+ lines across five log types (Audit, Puppet, SSHD, DHCP, Cron). On these complex files, \sys scales and yields high-precision/high-recall query results, especially when using LLM-derived semantic naming to create a custom schema. We provide step-wise microbenchmarks (template, schema, mapping), ablations, and efficiency evaluations. 

    \item \textbf{LogHub2.0~\cite{loghub} (\pref{ssec:LogHub2.0}).} LogHub2.0 is a benchmark of 14 log files used for template generation evaluation. While \sys is designed for end-to-end log parsing, we show our syntax parsing primitive is also suited for template generation. \sys equals or outperforms leading template generation methods, such as LILAC~\cite{jiang2024lilac}, Drain~\cite{Drain}, and Brain~\cite{brain23}, when cast to the canonical template-generation task.

    \item \textbf{Real-world logs (\pref{ssec:realworld}).} The real-world dataset contains 27 enterprise log types from Google SecOps with associated expert-written parsers. \sys generalizes to this dataset and delivers schema quality and retrieval results comparable to expert-written parsers without any manual effort.
\end{enumerate}

\noindent \textbf{Setup.} \sys is run with no human input using Gemini 2.5 Pro and the text-embedding-005 embedding model, unless otherwise stated.  Our source code, prompts, public SecurityLogs dataset, and curation tools to edit parsers and run queries are made public at 
\url{https://github.com/julien-piet/matryoshka/}. 
We rely on the Gemini~\cite{gemini} suite of models in our paper, but the code can support OpenAI models as well. We evaluate using Gemini 2.5 Flash and OpenAI o4-mini (see \cref{tab:metrics-query-ablation}) to confirm that the pipeline generalizes across model families. Our implementation supports both OCSF and UDM, but our mapping logic can be applied to any taxonomy. 

\begin{table}[t]
\centering
\caption{SecurityLogs parser summary}
\setlength{\tabcolsep}{3.5pt}
\begin{tabular}{|l|r|r|r|r|r|}
\hline
 & {\bf \begin{tabular}[b]{@{}l@{}} \small Total \\ \small nodes \end{tabular}} & 
   {\bf \begin{tabular}[b]{@{}l@{}} \small Variable \\ \small nodes \end{tabular}} & 
   {\bf \small Templates} & 
   {\bf \begin{tabular}[b]{@{}l@{}} \small Unique \\ \small fields \end{tabular}} & 
   {\bf \begin{tabular}[b]{@{}l@{}} \small Unique \\ \small OCSF \\ \small attributes \end{tabular}} \\
\hline
\hline
SSHD & 384 & 154 & 98 & 47 & 135  \\
\hline
Cron & 72 & 19 & 7 & 8 & 35\\
\hline
DHCP & 441 & 153 & 178 & 41 & 67\\
\hline
Audit & 6309 & 3000 & 496 & 150 & 286 \\
\hline
Puppet & 981 & 370 & 254 & 101 & 77 \\
\hline
\hline
\rowcolor[gray]{0.9} Total & 8187 & 3696 & 1033 & 347 & 600 \\
\hline
\end{tabular}
\label{tab:dataset_stats_bug_tracker}
\end{table}

\subsection{SecurityLogs}
\label{ssec:bugtracker}

SecurityLogs contains Linux system logs uploaded to a bug tracking website. These large and heterogeneous logs demonstrate \sys's ability to scale to formats with hundreds of different templates. Queries against our created schema consistently achieve high precision/recall and outperform naive substring matching. We additionally run ablation studies and introduce step-wise metrics to measure the performance of each stage of the parsing pipeline.

\subsubsection{Dataset}
We curated our dataset from Redhat Bugzilla bug reports~\cite{RedHatBugzilla}. We discovered that many users attach system log files to public bug reports there, so we crawled the Bugzilla website and downloaded all log files attached to bug reports up to August 2023. In total, this dataset contains over 30 million log lines. We filter to logs from five applications that are security-relevant and are well-enough documented that we could manually construct ground-truth parsers:

\begin{itemize}[leftmargin=*]
    \item \textbf{Linux kernel Audit logs (77K lines)}: Fine-grained logs of security-relevant events in Linux systems.
    \item \textbf{Puppet logs (157K lines)}: Logs from Puppet, an orchestration tool for server configuration deployment.
    \item \textbf{SSH daemon logs (35K lines)}: Logs involving SSH authentications and connections.
    \item \textbf{DHCP logs (378K lines)}: DHCP client logs reporting DHCP requests and lease information.
    \item \textbf{Cron logs (13K lines)}: Reports of scheduled Cron jobs.
\end{itemize}

\noindent
We ran \sys on each log file.
We created ground-truth parsers, by manually checking and fixing every template, field name, and mapping generated by \sys. The ground truth parsers are summarized in~\cref{tab:dataset_stats_bug_tracker}.

\subsubsection{Query comparison}
\label{ssec:bugtracker-query}
Our first evaluation measures \sys's ability to create queryable schemas for SecurityLogs.
We define 10 \emph{security-relevant queries} per log file that correspond 
to tasks in the NICCS NICE framework~\cite{niccs_nice_framework}, execute the queries, and
compare the resulting set of results to the ground-truth golden result, reporting
precision and recall.\footnote{We only wrote 5 
queries for Cron, which is less diverse than the four other log files.} 
We use our hand-written parsers to collect ground-truth answers. 
Each query is written in three variants:
\begin{enumerate}[leftmargin=*]
  \item a \textbf{standardized form} that uses canonical field names
        exactly as defined in the OCSF taxonomy;
  \item a \textbf{custom form} that uses the semantic field names
        produced by \sys for that log file. %
  \item a \textbf{naive substring form} that uses simple substring matches directly on raw log messages. 
\end{enumerate}

\begin{table}[t]
\centering
\small
\caption{\sys query precision and recall metrics}
\begin{tabular}{|l|c|c|c|c|c|c|}
\hline
\multirow{2}{*}{\textbf{Dataset}} & \multicolumn{2}{c|}{\textbf{OCSF}} & \multicolumn{2}{c|}{\textbf{Created}} & \multicolumn{2}{c|}{\textbf{Naive}} \\
& \multicolumn{2}{c|}{\textbf{attributes}} & \multicolumn{2}{c|}{\textbf{attributes}} & \multicolumn{2}{c|}{\textbf{substring}} \\
\cline{2-7}
 & \textbf{\small Prec.} & \textbf{\small Rec.} & \textbf{\small Prec.} & \textbf{\small Rec.} & \textbf{\small Prec.} & \textbf{\small Rec.} \\
\hline
\hline
SSHD & 0.90 & 0.90 & {\bf 1.00} & {\bf 1.00} & 0.90 & 0.80 \\
\hline
Cron & {\bf 1.00} & {\bf 1.00} & {\bf 1.00} & {\bf 1.00} & 0.92 & {\bf 1.00} \\
\hline
DHCP & 0.80 & 0.77 & {\bf 1.00} & {\bf 0.97} & 0.92 & 0.71 \\
\hline
Audit & 0.70 & 0.69 & {\bf 1.00} & {\bf 0.99} & 0.97 & 0.85 \\
\hline
Puppet & 0.90 & 0.88 & {\bf 1.00} & {\bf 0.98} & {\bf 1.00} & 0.63 \\
\hline
\hline
\rowcolor[gray]{0.9} Average & 0.86 & 0.85 & {\bf 1.00} & {\bf 0.99} & 0.94 & 0.80 \\
\hline
\end{tabular}
\label{tab:metrics-query}
\end{table}

\noindent
For example, one of our Cron log queries looks for scheduled jobs from a particular 
host within a specific time window, filtering on OCSF fields \texttt{time\_dt} 
and \texttt{device.hostname}, and checking for the presence of the \texttt{job.cmd\_line} 
OCSF field. In order to run expressive queries, we sometime include custom fields in 
standardized queries when one of the predicates was over a field that does not map to 
any OCSF attribute.

Standardized and custom queries are formed by applying unions or intersections of predicates. 
Predicates can be over the values of the fields, or over the static part of templates. 
Predicates on the static template are defined by substring matching. 
This limits the range of queries we can run. For example, ``return all lines indicating 
a bind error due to an already-in-use address'' is tedious to formulate in our query 
syntax, because it requires knowing all possible templates that could indicate such 
an event. A dedicated LLM-powered query planner could plausibly map queries to the relevant 
set of templates; we leave that to future work.

Substring matching is a common technique used by analysts to 
query unstructured log data. This technique is limited to simple queries (e.g., matching fixed phrases) and fails to express constraints on structured fields, time ranges, or events that occur in many different log formats.
In our experience, substring-match queries take longer to write than our structured queries. 
We assume an analyst will not have resources to comprehensively identify all message formats/templates that might have relevant information; instead, we simulate an analyst who looks for the first example they can find of a log message containing the necessary information and identifies a single substring from that log message to search for. We coin these queries \emph{naive} substring matches.



\smallskip \noindent \textbf{Results.}
The average precision and recall of the queries is reported in~\cref{tab:metrics-query}. We observe that using OCSF attributes for querying performs poorly compared to \sys-created attributes: This is due to the difficulties discussed in~\cref{ssec:security_operations}. If the mapping for a high-volume variable fails, this variable cannot be queried using OCSF attributes. We show in~\cref{ssec:realworld} that even expert-written parsers struggle at standardization: \sys-generated parsers slightly outperform expert-written ones on standardized queries.

In contrast, custom-form queries have perfect precision and near-perfect recall. 
The main error source is inconsistent naming of fields. For example, the Audit log has a field for the current working directory of a process. This field in the generated parser is most often called ``\texttt{current\_working\_directory}'', but sometimes abbreviated as ``\texttt{cwd}''. For the sake of simplicity, we chose to run queries on only one created field name, even when there are multiple names for the same concept: requiring analysts to list all possible fields would be tedious. Using an LLM query engine could help here, as the model could automatically select all relevant fields.


Queries over \sys-created field names consistently perform better than naive substring queries. OCSF-based querying achieves a similar F1 score to substring matching (0.85 for OCSF vs 0.86 for substring matching). OCSF queries have worse precision, due to different source fields collapsing into a single OCSF attribute, but better recall. Naive substring matching uses an arbitrary line as an example to model the substrings needed to run the query: this leads to missed lines when there are multiple possible templates that would match a given query, and to over-capture if the same substring is present in other lines. The complete set of substrings used for this task are detailed on our github, examples for DHCP are given in~\cref{app:queries}.

\begin{table}[t]
\centering
\caption{Average query precision/recall across ablations}
\begin{tabular}{|l|r|c|c|}
\hline
\textbf{Model} & \textbf{Ablation} & \textbf{OCSF} & \textbf{Custom} \\
\hline\hline
Gemini 2.5 Flash~\cite{gemini}      & none          & 0.76 / 0.76 & 0.92 / 0.91 \\
\hline
o4-mini~\cite{openai-o4mini-2025}   & none          & 0.83 / 0.81 & 0.98 / 0.95 \\
\hline
\multirow{3}{*}{Gemini 2.5 Pro~\cite{gemini}} 
                                    & none         & {\bf 0.86} / {\bf 0.85} & {\bf 1.00} / {\bf 0.99} \\
\cline{2-4}
                                    & \textit{no validation} 
                                                      & 0.79 / 0.77 & 0.99 / 0.96 \\
\cline{2-4}
                                    & \textit{no few-shot} 
                                                      & 0.42 / 0.43 & 0.74 / 0.71 \\
\hline
\end{tabular}
\label{tab:metrics-query-ablation}
\end{table}

\subsubsection{Ablations}
\label{ssec:ablations}
Our next experiments analyze how sensitive \sys is to different language models.
We also isolate the impact of \sys's validation and few-shot mechanisms to show parser consistency is not an emergent property of large model but the result of an architecture explicitly designed to build reliable parsers from unreliable language model outputs. We report average metrics in~\cref{tab:metrics-query-ablation}, leaving per-file results to~\cref{tab:metrics-query-ablation-full} in Appendix~\ref{app:add_results}.

\smallskip\noindent\textbf{Language model generalization.} 
We ran our pipeline using three language models: Gemini 2.5 Pro, Gemini 2.5 Flash\footnote{We still rely on Pro for the validation step, as it require advanced reasoning to be able to function.}, and OpenAI's o4-mini~\cite{openai-o4mini-2025}. 
Overall, Gemini 2.5 Flash performs worse than Pro, both using created and OCSF queries, most notably on Puppet logs. The model under-parsed Puppet log lines, keeping long error messages as single variables, instead of breaking them down into more granular component, which negatively impacts queryability. We recommend using advanced reasoning models with \sys. The prompts used in \sys were developed for the Gemini suite of models. o4-mini's performance (close to Gemini 2.5 Pro) shows the prompts generalize to other models, albeit with a slight decrease in performance. 

\smallskip\noindent\textbf{Architecture.}
We ran two additional experiments, one removing \sys's validation steps, and one without providing the model with few-shot examples of previous generations. Validation improves custom query recall, by coalescing variables that have different names into a single attribute, and also improves mapping significantly. Few-shot prompting helps the model maintain consistent syntax and semantics. Without few-shot examples, the model fails to parse similar structures in the same manner. The resulting parser is fragmented and queries achieve lower precision and recall.

\begin{table}[b]
\centering
\caption{Average query precision/recall across different architectures (all relying on Gemini-2.5-Pro).}
\begin{tabular}{|l|c|c|c|}
\hline
\textbf{Variant} & \textbf{OCSF} & \textbf{Custom} \\
\hline\hline
LILAC~\cite{jiang2024lilac} + naming heuristic & 0.24 / 0.21 & 0.42 / 0.31 \\
\hline
LILAC + \sys semantic steps & 0.34 / 0.31 & 0.60 / 0.60 \\
\hline
\sys & {\bf 0.86} / {\bf 0.85} & {\bf 1.00} / {\bf 0.99} \\
\hline
\end{tabular}
\label{tab:metrics-lilac-gap}
\end{table}

\subsubsection{Template generation shortcomings}
\label{ssec:gap-to-prior-work}

Prior work studies the problem of \emph{template generation}, whose goal is to associate each log entry with a wildcard template. While this representation provides useful signals for clustering and event recognition, it lacks the semantic information necessary for querying and advanced analytics. Moreover, we find that such representations are not suitable for log parsing, even when combined with a post-hoc semantic naming pipeline. To underscore this distinction in capabilities, we evaluate LILAC~\cite{jiang2024lilac}, a state-of-the-art template generation algorithm, as the syntax parsing layer of increasingly more complex end-to-end parsing pipelines:

\begin{itemize}[leftmargin=*]
    \item \textbf{LILAC + naming heuristic:} Generate wildcard templates using LILAC, then apply a single-prompt heuristic to associate each template with a schema and a second prompt to map schema fields to OCSF. Both prompts use fixed few-shot examples, provided in our source code.
    \item \textbf{LILAC + \sys semantics:} Convert LILAC-generated templates into the tree structure used by \sys, and run \sys's semantic stages (schema creation and mapping) on the resulting tree.
    \item \textbf{\sys end-to-end:} Use \sys for all three stages—syntax parsing, schema creation, and mapping.
\end{itemize}

All architectures rely on Gemini-2.5-Pro for parity. We compute query precision and recall as defined in~\cref{ssec:bugtracker-query}, and show results in~\cref{tab:metrics-lilac-gap}. LILAC with a naming heuristic achieves a precision of 0.42 and recall of 0.31; LILAC paired with \sys's semantic steps improves to 0.60 and 0.60; while the full \sys pipeline reaches 0.99 and 0.96. These results clearly illustrate the gap between template generation and end-to-end log parsing.

Upon inspection of these results, two factors dominate the poor performance of the LILAC-based architectures:
\begin{itemize}[leftmargin=*]
\item \textbf{Inconsistent naming:} Independently naming variables often leads to multiple names for the same concept. In the SSHD log, this approach produced three distinct field names for the source IP of an SSH connection: `\texttt{source\_IP}', `\texttt{client\_IP}', and `\texttt{ip\_address}'. This inconsistency complicates querying. \sys mitigates this through its parsing tree, which shares variables across templates, reuses previously named examples, and employs batched validation to ensure that identical concepts have a single canonical name.
\item \textbf{Hallucinated OCSF attributes:} The naive mapping stage often invents attributes that do not exist in the target taxonomy, reducing query precision. \sys prevents this by providing the model with a shortlist of valid fields and enforcing that any predicted attribute must be valid in the target taxonomy. Validation further ensures consistency across all templates. 
\end{itemize}

While adding \sys's semantic steps to LILAC improves both precision and recall, the resulting parsers still fall short of the full \sys pipeline. The remaining errors are not due to flaws in LILAC's algorithm, but rather to the intrinsic limitations of the wildcard representation. These limitations prevent accurate recovery of syntactic boundaries and constants, showing that the template generation task alone is insufficient to produce high-quality parsers. We attribute the gap to three root causes:

\begin{itemize}[leftmargin=*]
    \item \textbf{Blurry variable boundaries:} Wildcards set imprecise variable boundaries, which leads both to blurred fields and to over-capture. When two fields are adjacent, such as the date and hostname in `\texttt{Mar 9 23:46:29 puma25 sshd[17376]}', a template like `\texttt{<*> <*> sshd[<*>]}' cannot tell where one field ends and the next begins. This imprecision can lead to over-capture and collapse distinct log variants into a single template. For example, `\texttt{Mar 9 23:46:29 puma25 sshd[17376]: Accepted conn user:root pw}' may be matched by `\texttt{<*> <*> sshd[<*>: Accepted conn user:<*> <*>]}', which also matches the line in~\cref{fig:design-figure} but captures far more than intended. In contrast, \sys uses per-variable regular expressions to enforce precise boundaries and prevent such unintended capture.
    \item \textbf{Missing constants:} Wildcard templates omit constant tokens, leaving fixed values such as `\texttt{sshd}' unparsed and unqueryable. \sys instead isolates constants as distinct tokens that can be named, mapped, and queried.
\end{itemize}

Overall, these results yields two key takeaways. First, adding semantics to logs requires systematic validation to ensure consistent naming and accurate mapping; simply prompting an LLM without structure leads to inconsistency. Second, generating semantic-aware log parsers is fundamentally different from generating log templates. Even when state-of-the-art template generators are augmented with semantic steps, the resulting outputs remain unsuitable for accurate querying due to conflated variables, unparsed constants, and over-capture.





\subsubsection{Micro benchmarks}
\label{ssec:microbenchmarks}

We now evaluate individual steps on the same SecurityLogs files: syntax parsing, schema creation, and mapping.

\smallskip\noindent\textbf{Syntax parsing.}
%
We evaluate syntax parsing using similar metrics to those used for template generation. Template generation evaluation rely on two metrics. \emph{Group Accuracy (GA)} measures whether log lines derived from the same format string are grouped together. \emph{Parser Accuracy (PA)} checks if the predicted template exactly matches the ground truth. Both have drawbacks: PA is overly strict (splitting a field such as ``IP:port'' into two variables yields a score of zero even though the result may be more useful), while GA penalizes under-capture and over-capture equally. In practice, over-capture is worse, since it conflates distinct fields. For example, the template ``\texttt{Accepted <*> from <*>}'' is meant to capture usernames, but also absorbs IP addresses when present, merging unrelated variables.

To address these issues, we introduce two refined metrics. \emph{Template Similarity (TS)} measures how close the predicted template is to the ground truth when all variables are replaced by ``\texttt{<*>}''. It is computed as $1 - \frac{\text{Levenshtein}(\text{Ground Truth}, \text{Predicted})}{\max(|\text{Ground Truth}|, |\text{Predicted}|)}$, 
so small differences such as splitting or merging variables only reduce the score slightly rather than to zero. \emph{Parser Group Similarity (PGS)} measures how well the parser keeps log lines that share the same format string together. Over-capture (merging different formats into one group) yields a score of $0$, while under-capture (splitting one true group into smaller ones) receives partial credit given by $|\text{Parser Group}| / |\text{Ground Truth Group}|$.

We compute both metrics for \sys using Gemini 2.5 Pro. We provide as a reference the 
metrics for wildcard templates generated by two prior works---LILAC~\cite{jiang2024lilac} 
and Brain~\cite{brain23}. \sys outperforms both other works on all five logs. 
This performance gap can be attributed to two main factors: wildcard template have a tendency to over-capture, leading to poor parser group similarity, and prior work lacks consistency when parsing parallel structures in different log lines, leading to poor template similarity. The other drawbacks of wildcard templates mentioned in~\cref{ssec:gap-to-prior-work} impact end-to-end performance but are not captured by these metrics. 

\begin{table}[t]
\centering
\caption{Syntax parsing metrics versus related work.}
\begin{tabular}{|l|r|c|c|c|c|c|c|}
\hline
\multirow{2}{*}{\textbf{Dataset}} & \multirow{2}{*}{\textbf{Lines}} & \multicolumn{2}{c|}{\textbf{\sys}} & \multicolumn{2}{c|}{\textbf{LILAC}} & \multicolumn{2}{c|}{\textbf{brain23}} \\
\cline{3-8}
& & \textbf{PGS} & \textbf{TS} & \textbf{PGS} & \textbf{TS} & \textbf{PGS} & \textbf{TS} \\
\hline
\hline
SSHD   & 35,329  & \textbf{1.00} & \textbf{0.90} & 0.95 & 0.78 & 0.88 & 0.73 \\
\hline
Cron   & 12,547  & \textbf{1.00} & \textbf{0.80} & \textbf{1.00} & \textbf{0.80} & \textbf{1.00} & 0.60 \\
\hline
DHCP   & 377,653 & \textbf{0.98} & \textbf{0.89} & 0.18 & 0.68 & 0.54 & 0.69 \\
\hline
Audit  & 76,636  & \textbf{0.97} & \textbf{0.98} & 0.26 & 0.78 & 0.62 & 0.52 \\
\hline
Puppet & 156,880 & \textbf{0.97} & \textbf{0.99} & 0.77 & 0.76 & 0.96 & 0.68 \\
\hline
\end{tabular}
\label{parser-results}
\end{table}

\begin{table}[b]
\centering
\caption{\sys schema and mapping metrics}
\begin{tabular}{|l|c|c|c|c|c|}
\hline
 & \textbf{SSHD} & \textbf{Cron} & \textbf{DHCP} & \textbf{Audit} & \textbf{Puppet} \\
\hline\hline
\makecell[l]{Schema Group\\Similarity} & 0.95 & 1.00 & 0.94 & 0.64 & 0.85 \\
\hline
Mapping Accuracy & 0.93 & 0.93 & 0.59 & 0.79 & 0.98 \\
\hline
\end{tabular}
\label{tab:metrics-schema-mapping}
\end{table}

\smallskip\noindent\textbf{Schema creation. }
Schema creation entails clustering variables by meaning (e.g., assigning every ``source IP'' variable to a field ``\texttt{source\_ip}''). For this step, we define \emph{Schema Group Similarity (SGS)}. We group variables by their assigned name and score each variable. If multiple ground truth fields are merged (overcapture), the score is 0. If a ground truth field is split into several fields (undercapture), the score is the ratio of the parser's group size to that of the ground-truth cluster. This rewards grouping variables of the same semantic role without merging separate roles. The final score is the weighted average of variable scores depending on their frequency. 

%
We ran \sys's schema creation on ground-truth templates so that both parsers expose the same set of variables. Results are reported in~\cref{tab:metrics-schema-mapping}. SGS is nearly perfect for SSHD, Cron, and DHCP. Puppet and Audit score lower at 0.85 and 0.64, because high-volume variables such as the ``\texttt{operation}'' field are split across multiple names. In rarer cases, the created schema slightly over-captures: ``\texttt{audit\_user\_id}'' is merged ``\texttt{new\_audit\_user\_id}''.

\smallskip\noindent\textbf{Mapping.}
The final step evaluates how well fields are mapped to standardized attributes. We define a per-field accuracy score: if a parser assigns no mapping to a field that also has no mapping in the ground truth, it is counted correct. Otherwise, accuracy is the fraction of assigned attributes that also appear in the ground-truth set. The mapping accuracy is the weighted average of the scores. 

%
We ran mapping independently on ground truth schemas, using Gemini 2.5 Pro, and restricted the number of mappings to at most one per source attribute. Results are shown in~\cref{tab:metrics-schema-mapping}. SSHD, Puppet and Cron mappings are mostly correct. Audit and DHCP score lower (at 0.79 and 0.59), due to mistakes on high volume variables. The logging server's hostname and assigned IP map to OCSF's ``\texttt{device.hostname}`` and ``\texttt{device.ip}``, respectively. The generated parser assigns these to attributes in the ``\texttt{src\_endpoint}`` tree, which is less appropriate.

\subsubsection{Time and cost evaluation}
\begin{table}[t]
\centering
\caption{Timing and LLM token usage (in/out in millions of tokens) for \sys steps}
\begin{tabular}{|l|r|r|r|r|r|}
\hline
{\bf Step} & {\bf SSHD} & {\bf Cron} & {\bf DHCP} & {\bf Audit} & {\bf Puppet} \\
\hline
\hline
Generation & 1h 45m & 28m & 4h 16m & 10h 7m & 4h 16m \\
\hline
Creation & 30m & 2m & 2h 16m & 9h 33m & 1h 13m \\
\hline
Mapping & 1h 7m & 9m & 1h 24m & 4h 30m & 2h 13m \\
\hline
\hline
\rowcolor[gray]{0.9} Total & 3h 22m & 39m & 7h 56m & 24h 10m & 7h 42m \\
\hline
\rowcolor[gray]{0.95} MTok In/Out & 21 / 3 & 2 / 1 & 33 / 3 & 162 / 15 & 50 / 5 \\
\hline
\end{tabular}
\label{tab:timing_summary}
\end{table}

Our system allows running queries in a few seconds, even when the source log file is hundreds of thousands of lines. Log ingestion is also fast: we consistently parse log files at over 500 lines per second. This is possible because LLMs are only used at generation time, not at run-time: live data is statically parsed.
However, the process of generating parsers is slow, due to the fact most steps use previous answers of the model to few-shot prompt the next ones, thus cannot be parallelized. \sys takes on average 150 seconds per template. Run times and token usage are detailed in~\cref{tab:timing_summary}. \sys only needs to be run once, but future work should look at speeding up this process.  We estimate the number of input and output tokens required for each file---system prompts and repeated queries are cached to reduce costs. Averaging over the five files, each template requires about 250K input tokens and 25K output tokens (including every step's generation and validation), or an average of 0.60 USD per template when using Gemini 2.5 Pro or 0.14 USD with Gemini 2.5 Flash. As a back-of-the-envelope comparison, Google SecOps pushes 200 parser updates monthly, requiring 4,000 engineer-hours, or 20 hours per update. A typical update touches a handful of templates; conservatively assuming $\sim$10 templates per update yields 2 hours per template (time spent researching field semantics, selecting mappings, and writing code). \sys is $40\times$ faster and $25\times$ cheaper at U.S. federal minimum wage when using Gemini 2.5 Pro.


\subsection{LogHub2.0}
\label{ssec:LogHub2.0}

The second dataset we evaluate against is an existing benchmark for template generation works. While prior work's wildcard templates are insufficient for end-to-end parsing (\cref{ssec:gap-to-prior-work}), templates generated as part of \sys's syntax parser can be converted back to wildcard templates, which equal or outperform those created by state of the art approaches such as LILAC~\cite{jiang2024lilac}, Drain~\cite{Drain}, and Brain~\cite{brain23} when evaluated under the same settings and language model.

\subsubsection{Dataset}

Existing approaches to template generation commonly use LogHub2.0~\cite{jiang2024, loghub}. This resource contains 14 log files (over 3 million lines) annotated with ground-truth wildcard-based templates. We use it to compare the performance of our template generation step against prior methods, but we do not generate ground-truth labels for schema creation or attribute mapping because (1) most of the data is outdated (some logs are over 20 years old); (2) several files are heavily anonymized, limiting reliable semantic extraction; and (3) most are not security-focused and instead are activity logs from supercomputers or distributed systems.


\subsubsection{Experiment} 
We selected three prior works with the best performance to compare against: 
LILAC~\cite{jiang2024lilac}, 
one of the most promising LLM-based approaches, Brain~\cite{brain23}, a lightweight 
parser, and Drain3, the latest version of Drain~\cite{Drain}, a popular log parsing 
algorithm. On the LogHub2.0 dataset, these three frameworks use a pre-parsed version of
the logs and only generate templates for the suffixes. Drain and Brain do not use LLMs. 
To ensure an equal comparison, both LILAC and \sys are run using Gemini 2.5 Flash, using a single human-labeled example (LILAC used human-labeled examples in their evaluation).
\smallskip \noindent {\bf Results.}
We report two metrics, defined in~\cref{ssec:microbenchmarks}: parser group similarity, measuring whether log lines derived from the same format string are grouped together (PGS), 
and template similarity, measuring how close each template is to the ground truth (TS). Full results are in~\cref{fig:LogHub2.0-full} in Appendix~\ref{app:add_results}. \sys matches or outperforms other works on most log files. 
On average, we obtain a parser group similarity of 0.97 and a template similarity of 0.92, while the best of the other schemes, 
LILAC, obtains a PGS of 0.95 and a TS of 0.87. Drain3 achieves a PGS=0.89 and TS=0.81, while Brain gets PGS=0.86 and TS=0.74. 
When using the traditional metrics (parser accuracy and group accuracy), \sys improves on LILAC's parser accuracy, getting 0.70 
instead of 0.63, but is worse for group accuracy, with 0.85 instead of 0.88. This is reflective of \sys's tendency to parse 
variables with more granularity. For example, LogHub2.0 provides the following template in the Spark log file: 
`\texttt{Error sending message [message = <*>] in <*> attempts}'. \sys further split this 
into four different templates based on the content of the message, with sub-templates such as `\texttt{message = GetLocations(<*>)}' or 
`\texttt{message = RetrieveSparkProps}'. In contrast, LILAC generated a single template, `\texttt{Error sending message [<*>] in <*> attempts}'.
The performance gap is not as significant as observed in~\cref{ssec:microbenchmarks}, because in the LogHub2.0 dataset similar prefixes across 
lines are pre-parsed, making consistency less important.

\subsection{Real-world evaluation}
\label{ssec:realworld}

Our last dataset comprises real-world enterprise log files. We leverage this data to evaluate \sys on diverse log types, and show we rival expert-written parsers while being fully automated and requiring no manual effort.

\subsubsection{Dataset}
\label{ssec:enterprise_data}
Our dataset comprises real-world log samples from 27 diverse log types with associated human-written parsers from Google SecOps. These include web-server access logs, intrusion-prevention alerts, user-management audits, and endpoint-protection telemetry.  Each sample is modest in size---no more than a few hundred lines---but collectively they span a spectrum of events found in real-world enterprise environments. We use this corpus to assess \sys\ on the long-tail of real-world log types and to compare its output against the hand-written parsers.

\subsubsection{Side-by-side holistic comparison}
\label{ssec:sidebyside}
We ingest logs into structured UDM format using two parsers---an expert-written parser, and a \sys-generated one. We compare both these schemas along four axes: \emph{coverage} (fraction of source fields mapped), \emph{accuracy} (semantic correctness of each mapping), \emph{consistency} (stability of mappings across a line), and \emph{queryability} (how easily an analyst can express predicates).

\smallskip\noindent\textbf{Step 1: Expert assessment.}
We drew a random sample of log lines along with both schemas, presented them blindly to an experienced UDM rule writer, 
and asked for a 1--5 preference score  
(1/5 = strong preference for the left/right schema, 2/4 = weak preference, 3 = tie) using the four axes above.  
Across 15 comparisons the expert judged the schemas \textit{equivalent} in 8 cases, preferred \sys in 3, and preferred the UDM baseline in 4. While this sample is modest---reflecting limited expert availability---we use it primarily to calibrate our LLM autorater.

\smallskip\noindent\textbf{Step 2: LLM autorater.}
We use the expert-labelled sample to calibrate a Gemini 2.0 Pro autorater---a model from a different family to those used during generation time---that compares the two schemas. For each log line, the model receives the raw text, the Matryoshka fields, the expert-parser fields, and a four-axis rubric, then outputs a 1–5 preference score. We randomize parser order, sample multiple completions, and average scores to eliminate positional bias. We retain the prompt variant that exactly reproduces the expert’s preferences on the calibration set.

\smallskip\noindent\textbf{Results.}
After calibration the autorater graded every line in the 27 files.  
It preferred \sys\ in 25\% of cases, favored the UDM baseline 
in 24\%, and reported no preference in 49\%.  
These findings echo the expert's conclusion: \sys\ delivers schema 
quality comparable to expert-written parsers---without any manual work. 
We substantiate these observations with larger-scale quantitative results in the subsequent subsection.

\subsubsection{Query comparison}
\label{ssec:e2e-def}
We further compare \sys's UDM output to expert-written UDM parsers by measuring how well \sys parsers support queries. Similar to~\cref{ssec:bugtracker-query}, we define security-relevant queries, execute them and compare the results to a manually derived ground-truth response set.

\smallskip\noindent\textbf{Experimental setup.}
We select 10 files at random from the 27, devise a security-relevant query for each, 
informed by NICE tasks, and manually identify the correct matches. The queries span tasks such as account monitoring, network reconnaissance, or resource-access analysis.
Then, each query is expressed in a standard UDM form, and in a custom-schema form 
using \sys-generated field names. The UDM query is written agnostically of the parser, using the taxonomy's documentation. We then run three configurations: (1) expert-written UDM parser, queried with the UDM-form query; (2) \sys-generated UDM parser, queried with the UDM-form query; (3) \sys-generated parser, queried with the custom-schema form query.

\begin{table}[t]
\centering
\small
\caption{Enterprise-log query accuracy}
\begin{tabular}{|l|c|c|}
\hline
\textbf{Configuration} & \textbf{Precision} & \textbf{Recall} \\
\hline\hline
Expert-written UDM parser        & 0.50 & 0.48 \\
\hline
\sys\ + UDM query                   & 0.60 & 0.60 \\
\hline
\sys\ + custom-schema query         & {\bf 1.00} & {\bf 0.95} \\
\hline
\end{tabular}
\label{tab:enterprise_udm}
\end{table}

\smallskip\noindent\textbf{Results.} Custom-schema queries yield near-perfect performance, while both UDM parsers struggle. Three points stand out: \textit{(i)} Queries against taxonomies are not accurate, even when using expert-written parsers. This is representative of the mapping difficulties discussed in~\cref{ssec:security_operations}. \textit{(ii)} \sys UDM parsers are less accurate (fields are not always mapped to the best candidate attribute) than expert-written UDM parsers, but have better coverage. \textit{(iii)} Queries using \sys's custom schema yield far more precise results, demonstrating that log-specific field names enable more accurate predicates, albeit at the cost of learning a file-local vocabulary rather than a general-purpose, normalized schema.

\section{Discussion}
\label{discussion}

\subsection{To normalize or not to normalize}
\label{ssec:query_discussion}

Our evaluations show mapping logs to standard taxonomies is difficult,
even for expert analysts. Misalignment between the specificity of source 
logs and taxonomies both under-capture domain-specific 
concepts and provide excessive nuance for standard concepts 
(\pref{ssec:security_operations}). Despite these issues, industry continues to use 
common taxonomies for normalization. They allow for detection rules that are agnostic to the source data, and enable cross-correlation in ways custom schemas cannot. 
Yet, we find custom schemas (i.e., semantic naming) allow for more precise search over logs, since they are created on a per-log basis. We suggest that future research could explore parsing logs to a custom schema, and using LLMs to compile a natural language query/rule to custom-schema fields, and skip normalization entirely.

\subsection{Limitations}
\label{ssec:limitations}

Although~\sys advances end-to-end log parsing, several limitations remain. 
As noted above, mapping extracted fields to standard schema attributes is challenging due 
to the volume of target attributes and nuances in field definitions. Queries against these incorrect or missing mappings can lead to missed attacks or false positives. Given analysts' need for high reliability, any incorrect or missing fields can significantly impact usefulness, though as we have seen this is a common problem even for human-generated parsers. To mitigate this, we built a prototype 
interface for analysts to inspect and correct parsers generated by \sys, though automated methods to identify and fix these errors based on query-time usage remain an interesting area of potential future work.
\section{Conclusion}
\label{conclusion}

We presented \sys, the first end-to-end system leveraging LLMs to automatically generate semantically-aware structured log parsers. \sys combines a novel syntactic parser with a semantic layer that clusters variables, maps them to structured schemas, assigns contextually meaningful field names, and maps variables to attributes from standard taxonomies. 
While log standardization remains challenging, \sys parsers rival expert-written parsers on real-world logs both in qualitative head-to-head comparisons and in query accuracy, removing much of the costly human effort needed to craft parsers. \sys-created schemas offer an alternative to standardized taxonomies that enable precise querying ($F_1=0.99$), significantly higher than achievable with existing substring-matching techniques, at a similar cost in query writing. By automatically transforming unstructured logs into structured, semantically rich data, \sys represents a meaningful step toward enabling security analysts to focus on threat detection rather than manual parser construction. 

\section*{Acknowledgments}
We thank Sizhe Chen for gathering the logs from the Redhat Bugzilla. We thank Aashish Sharma and the cybersecurity team at Lawrence Berkeley National Laboratory for providing valuable insights into security operations and helping shape this project. We also thank Calvin Kim for his expert advice in comparing \sys to existing UDM parsers, Sunil Vasisht for his insightful feedback that helped shape \sys, and David Huang for helping implement template generation works. 
This research was supported by the KACST-UCB Joint Center on Cybersecurity, OpenAI, the National Science Foundation under grant numbers 2229876 (the ACTION center) and CNS-2154873, the Department of Homeland Security, IBM, C3.ai Digital Transformation Institute, Open Philanthropy, and Google. Any opinions, findings, and conclusions or recommendations expressed in this material are those of the author(s) and do not necessarily reflect the views of the sponsors.

\bibliographystyle{IEEEtran}
\bibliography{refs}

@inproceedings{smat,
author = {Zhang, Jing and Shin, Bonggun and Choi, Jinho D. and Ho, Joyce C.},
title = {{SMAT}: An Attention-Based Deep Learning Solution to the Automation of Schema Matching},
year = {2021},
booktitle = {ADBIS},
}

@article{rahm2001survey,
  title = {A survey of approaches to automatic schema matching},
  author={Erhard Rahm and Philip A. Bernstein},
  journal = {VLDB},
  year = {2001},
}

@article{hong2024next,
  title={Next-generation database interfaces: A survey of {LLM}-based text-to-{SQL}},
  author={Hong, Zijin and Yuan, Zheng and Zhang, Qinggang and Chen, Hao and Dong, Junnan and Huang, Feiran and Huang, Xiao},
  journal={arXiv preprint arXiv:2406.08426},
  year={2024}
}

@inproceedings{rag,
author = {Lewis, Patrick and Perez, Ethan and Piktus, Aleksandra and Petroni, Fabio and Karpukhin, Vladimir and Goyal, Naman and K\"{u}ttler, Heinrich and Lewis, Mike and Yih, Wen-tau and Rockt\"{a}schel, Tim and Riedel, Sebastian and Kiela, Douwe},
title = {Retrieval-augmented generation for knowledge-intensive {NLP} tasks},
year = {2020},
booktitle = {NeurIPS},
}

@article{karlsen2024,
author = {Karlsen, Egil and Luo, Xiao and Zincir-Heywood, Nur and Heywood, Malcolm},
title = {Benchmarking Large Language Models for Log Analysis, Security, and Interpretation},
year = {2024},
journal = {Journal of Network and Systems Management},
}

@misc{dai2024uqequeryengineunstructured,
      title={{UQE}: A Query Engine for Unstructured Databases}, 
      author={Hanjun Dai and Bethany Yixin Wang and Xingchen Wan and Bo Dai and Sherry Yang and Azade Nova and Pengcheng Yin and Phitchaya Mangpo Phothilimthana and Charles Sutton and Dale Schuurmans},
      year={2024},
      eprint={2407.09522},
      archivePrefix={arXiv},
      primaryClass={cs.DB},
}

@misc{sheetrit2024rematchretrievalenhancedschema,
      title={{ReMatch}: Retrieval Enhanced Schema Matching with LLMs}, 
      author={Eitam Sheetrit and Menachem Brief and Moshik Mishaeli and Oren Elisha},
      year={2024},
      eprint={2403.01567},
      archivePrefix={arXiv},
      primaryClass={cs.DB},
      primaryClass={cs.DB},
}

@misc{parciak2024schemamatchinglargelanguage,
      title={Schema Matching with Large Language Models: an Experimental Study}, 
      author={Marcel Parciak and Brecht Vandevoort and Frank Neven and Liesbet M. Peeters and Stijn Vansummeren},
      year={2024},
      eprint={2407.11852},
      archivePrefix={arXiv},
      primaryClass={cs.DB},
}

@misc{zhang2024smutfschemamatchingusing,
      title={{SMUTF}: Schema Matching Using Generative Tags and Hybrid Features}, 
      author={Yu Zhang and Mei Di and Haozheng Luo and Chenwei Xu and Richard Tzong-Han Tsai},
      year={2024},
      eprint={2402.01685},
      archivePrefix={arXiv},
      primaryClass={cs.CL},
}

@misc{liu2024magnetocombiningsmalllarge,
      title={Magneto: Combining Small and Large Language Models for Schema Matching}, 
      author={Yurong Liu and Eduardo Pena and Aecio Santos and Eden Wu and Juliana Freire},
      year={2024},
      eprint={2412.08194},
      archivePrefix={arXiv},
      primaryClass={cs.DB},
}

@misc{xu2025kcmfknowledgecompliantframeworkschema,
      title={{KcMF}: A Knowledge-compliant Framework for Schema and Entity Matching with Fine-tuning-free LLMs}, 
      author={Yongqin Xu and Huan Li and Ke Chen and Lidan Shou},
      year={2025},
      eprint={2410.12480},
      archivePrefix={arXiv},
      primaryClass={cs.CL},
}

@inproceedings{fu2009execution,
  title={Execution anomaly detection in distributed systems through unstructured log analysis},
  author={Fu, Qiang and Lou, Jian-Guang and Wang, Yi and Li, Jiang},
  booktitle={ICDM},
  year={2009},
}

@inproceedings{tang2011logsig,
  title={{LogSig}: Generating system events from raw textual logs},
  author={Tang, Liang and Li, Tao and Perng, Chang-Shing},
  booktitle={CIKM},
  year={2011}
}

@inproceedings{hamooni2016logmine,
  title={{LogMine}: Fast pattern recognition for log analytics},
  author={Hamooni, Hossein and Debnath, Biplob and Xu, Jianwu and Zhang, Hui and Jiang, Guofei and Mueen, Abdullah},
  booktitle={CIKM},
  year={2016}
}

@article{brain23,
  author={Yu, Siyu and He, Pinjia and Chen, Ningjiang and Wu, Yifan},
  journal={IEEE TSC}, 
  title={Brain: Log Parsing With Bidirectional Parallel Tree}, 
  year={2023},
}

@inproceedings{IPLoM,
author = {Makanju, Adetokunbo A.O. and Zincir-Heywood, A. Nur and Milios, Evangelos E.},
title = {Clustering event logs using iterative partitioning},
year = {2009},
booktitle = {KDD},
}

@article{Spell,
  author={Du, Min and Li, Feifei},
  journal={IEEE TKDE}, 
  title={Spell: Online Streaming Parsing of Large Unstructured System Logs}, 
  year={2019},
}

@inproceedings{Drain,
  author={He, Pinjia and Zhu, Jieming and Zheng, Zibin and Lyu, Michael R.},
  booktitle={ICWS}, 
  title={Drain: An Online Log Parsing Approach with Fixed Depth Tree}, 
  year={2017},
}

@inproceedings{POP,
  author={He, Pinjia and Zhu, Jieming and He, Shilin and Li, Jian and Lyu, Michael R.},
  booktitle={IEEE TDSC}, 
  title={Towards Automated Log Parsing for Large-Scale Log Data Analysis}, 
  year={2018},
}

@inproceedings{SPINE,
author = {Wang, Xuheng and Zhang, Xu and Li, Liqun and He, Shilin and Zhang, Hongyu and Liu, Yudong and Zheng, Lingling and Kang, Yu and Lin, Qingwei and Dang, Yingnong and Rajmohan, Saravanakumar and Zhang, Dongmei},
title = {{SPINE}: a scalable log parser with feedback guidance},
year = {2022},
booktitle = {ESEC/FSE},
}

@inproceedings{mizutani2013incremental,
  title={Incremental mining of system log format},
  author={Mizutani, Masayoshi},
  booktitle={SCC},
  year={2013},
  organization={IEEE}
}

@misc{shima2016length,
  title={Length matters: Clustering system log messages using length of words},
  author={Shima, Keiichi},
  eprint={1611.03213},
  archivePrefix={arXiv},
  primaryClass={cs.OH},
  year={2016}
}

@inproceedings{vaarandi2003data,
  title={A data clustering algorithm for mining patterns from event logs},
  author={Vaarandi, Risto},
  booktitle={IPOM},
  year={2003},
}

@inproceedings{jiang2008abstracting,
  title={Abstracting execution logs to execution events for enterprise applications (short paper)},
  author={Jiang, Zhen Ming and Hassan, Ahmed E and Flora, Parminder and Hamann, Gilbert},
  booktitle={QSIC},
  year={2008},
}

@inproceedings{nagappan2010abstracting,
  title={Abstracting log lines to log event types for mining software system logs},
  author={Nagappan, Meiyappan and Vouk, Mladen A},
  booktitle={MSR},
  year={2010},
}

@inproceedings{vaarandi2015logcluster,
  title={{LogCluster} - A data clustering and pattern mining algorithm for event logs},
  author={Vaarandi, Risto and Pihelgas, Mauno},
  booktitle={CNSM},
  year={2015},
}

@article{dai2020logram,
  title={Logram: Efficient log parsing using $n$-gram dictionaries},
  author={Dai, Hetong and Li, Heng and Chen, Che-Shao and Shang, Weiyi and Chen, Tse-Hsun},
  journal={IEEE TSE},
  year={2020},
}

@inproceedings{huo2023semparser,
  title={{SemParser}: A semantic parser for log analytics},
  author={Huo, Yintong and Su, Yuxin and Lee, Cheryl and Lyu, Michael R},
  booktitle={ICSE},
  year={2023},
}

@misc{ocsf,
title={{OCSF Schema -- Categories}},
author={{Open Cybersecurity Schema Framework}},
year={2025},
note={\url{https://schema.ocsf.io/1.4.0/}}
}

@inproceedings{tak2021lognroll,
  title={Lognroll: Discovering accurate log templates by iterative filtering},
  author={Tak, Byungchul and Han, Wook-Shin},
  booktitle={Middleware},
  year={2021}
}

@inproceedings{yu2024loggenius,
  title={{LogGenius}: An Unsupervised Log Parsing Framework with Zero-shot Prompt Engineering},
  author={Yu, Xian and Nong, Shengxi and He, Dongbiao and Zheng, Weijie and Ma, Teng and Liu, Ning and Li, Jianhui and Xie, Gaogang},
  booktitle={ICWS},
  year={2024},
}

@misc{zhang2024lemur,
  title={Lemur: Log parsing with entropy sampling and chain-of-thought merging},
  author={Zhang, Wei and Guo, Hongcheng and Le, Anjie and Yang, Jian and Liu, Jiaheng and Li, Zhoujun},
  year={2025},
  eprint={2402.18205},
  archivePrefix={arXiv},
  primaryClass={cs.SE},
}

@inproceedings{xiao2024free,
  title={Demonstration-Free: Towards More Practical Log Parsing with Large Language Models},
  author={Xiao, Yi and Le, Van-Hoang and Zhang, Hongyu},
  booktitle={ASE},
  year={2024}
}

@misc{ma2024librelog,
  title={{LibreLog}: Accurate and Efficient Unsupervised Log Parsing Using Open-Source Large Language Models},
  author={Ma, Zeyang and Kim, Dong Jae and Chen, Tse-Hsun},
  eprint={2408.01585},
  archivePrefix={arXiv},
  primaryClass={cs.SE},
  year={2024}
}

@misc{ma2024luk,
  title={Luk: Empowering log understanding with expert knowledge from large language models},
  author={Ma, Lipeng and Yang, Weidong and Jiang, Sihang and Fei, Ben and Zhou, Mingjie and Li, Shuhao and Zhao, Mingyu and Xu, Bo and Xiao, Yanghua},
eprint={2409.01909},
      archivePrefix={arXiv},
      primaryClass={cs.SE},
  year={2024}
}

@inproceedings{zhong2024logparser,
  title={{LogParser-LLM}: Advancing efficient log parsing with large language models},
  author={Zhong, Aoxiao and Mo, Dengyao and Liu, Guiyang and Liu, Jinbu and Lu, Qingda and Zhou, Qi and Wu, Jiesheng and Li, Quanzheng and Wen, Qingsong},
  booktitle={KDD},
  year={2024}
}

@inproceedings{liu2022uniparser,
  title={Uniparser: A unified log parser for heterogeneous log data},
  author={Liu, Yudong and Zhang, Xu and He, Shilin and Zhang, Hongyu and Li, Liqun and Kang, Yu and Xu, Yong and Ma, Minghua and Lin, Qingwei and Dang, Yingnong and others},
  booktitle={WWW},
  year={2022}
}

@inproceedings{li2023did,
  title={Did we miss something important? studying and exploring variable-aware log abstraction},
  author={Li, Zhenhao and Luo, Chuan and Chen, Tse-Hsun and Shang, Weiyi and He, Shilin and Lin, Qingwei and Zhang, Dongmei},
  booktitle={ICSE},
  year={2023},
}

@inproceedings{le2023log,
  title={Log parsing with prompt-based few-shot learning},
  author={Le, Van-Hoang and Zhang, Hongyu},
  booktitle={ICSE},
  year={2023},
}

@inproceedings{xu2024divlog,
  title={{DivLog}: Log parsing with prompt enhanced in-context learning},
  author={Xu, Junjielong and Yang, Ruichun and Huo, Yintong and Zhang, Chengyu and He, Pinjia},
  booktitle={ICSE},
  year={2024}
}

@inproceedings{ma2024llmparser,
  title={{LLMParser}: An exploratory study on using large language models for log parsing},
  author={Ma, Zeyang and Chen, An Ran and Kim, Dong Jae and Chen, Tse-Hsun and Wang, Shaowei},
  booktitle={ICSE},
  year={2024}
}

@article{brown2020language,
  title={Language models are few-shot learners},
  author={Brown, Tom and Mann, Benjamin and Ryder, Nick and Subbiah, Melanie and Kaplan, Jared D and Dhariwal, Prafulla and Neelakantan, Arvind and Shyam, Pranav and Sastry, Girish and Askell, Amanda and others},
  journal={NeurIPS},
  year={2020}
}

@article{jiang2024lilac,
  title={{LILAC}: Log parsing using {LLM}s with adaptive parsing cache},
  author={Jiang, Zhihan and Liu, Jinyang and Chen, Zhuangbin and Li, Yichen and Huang, Junjie and Huo, Yintong and He, Pinjia and Gu, Jiazhen and Lyu, Michael R},
  journal={FSE},
  year={2024},
}

@inproceedings{jiang2024large,
  title={A large-scale evaluation for log parsing techniques: How far are we?},
  author={Jiang, Zhihan and Liu, Jinyang and Huang, Junjie and Li, Yichen and Huo, Yintong and Gu, Jiazhen and Chen, Zhuangbin and Zhu, Jieming and Lyu, Michael R},
  booktitle={ISSTA},
  year={2024}
}

@inproceedings{astekin2024comparative,
  title={A Comparative Study on Large Language Models for Log Parsing},
  author={Astekin, Merve and Hort, Max and Moonen, Leon},
  booktitle={ESEM},
  year={2024}
}

@misc{huang2024lunar,
  title={{LUNAR}: Unsupervised {LLM}-based Log Parsing},
  author={Huang, Junjie and Jiang, Zhihan and Chen, Zhuangbin and Lyu, Michael R},
eprint={2406.07174},
      archivePrefix={arXiv},
      primaryClass={cs.SE},
  year={2024}
}

@INPROCEEDINGS {loghub,
author = { Zhu, Jieming and He, Shilin and He, Pinjia and Liu, Jinyang and Lyu, Michael R. },
booktitle = {ISSRE},
title = {{ Loghub: A Large Collection of System Log Datasets for AI-driven Log Analytics }},
year = {2023},
}

@misc{chen2023teachinglargelanguagemodels,
      title={Teaching Large Language Models to Self-Debug}, 
      author={Xinyun Chen and Maxwell Lin and Nathanael Schärli and Denny Zhou},
      year={2023},
      eprint={2304.05128},
      archivePrefix={arXiv},
      primaryClass={cs.CL},
      url={https://arxiv.org/abs/2304.05128}, 
}

@inproceedings{jiang2024,
author = {Jiang, Zhihan and Liu, Jinyang and Huang, Junjie and Li, Yichen and Huo, Yintong and Gu, Jiazhen and Chen, Zhuangbin and Zhu, Jieming and Lyu, Michael R.},
title = {A Large-Scale Evaluation for Log Parsing Techniques: How Far Are We?},
year = {2024},
isbn = {9798400706127},
publisher = {Association for Computing Machinery},
address = {New York, NY, USA},
url = {https://doi.org/10.1145/3650212.3652123},
doi = {10.1145/3650212.3652123},
booktitle = {Proceedings of the 33rd ACM SIGSOFT International Symposium on Software Testing and Analysis},
pages = {223–234},
numpages = {12},
location = {Vienna, Austria},
series = {ISSTA 2024}
}

@software{splunk,
  title = {Splunk},
  author = {{Splunk Inc.}},
  year = {2025},
  url = {https://www.splunk.com},
  note = {Accessed: 2025-04-12}
}

@software{gso,
  title = {Google Security Operations},
  author = {{Google LLC}},
  year = {2025},
  url = {https://cloud.google.com/security/products/security-operations},
  note = {Accessed: 2025-04-12}
}

@software{elastic,
  title = {Elastic Security},
  author = {{ Elasticsearch B.V.}},
  year = {2025},
  url = {https://www.elastic.co/security},
  note = {Accessed: 2025-04-12}
}

@misc{liu2025optimizingllmqueriesrelational,
      title={Optimizing {LLM} Queries in Relational Data Analytics Workloads}, 
      author={Shu Liu and Asim Biswal and Amog Kamsetty and Audrey Cheng and Luis Gaspar Schroeder and Liana Patel and Shiyi Cao and Xiangxi Mo and Ion Stoica and Joseph E. Gonzalez and Matei Zaharia},
      year={2025},
      eprint={2403.05821},
      archivePrefix={arXiv},
      primaryClass={cs.LG},
      url={https://arxiv.org/abs/2403.05821}, 
}

@misc{anderson2024designllmpoweredunstructuredanalytics,
      title={The Design of an {LLM}-powered Unstructured Analytics System}, 
      author={Eric Anderson and Jonathan Fritz and Austin Lee and Bohou Li and Mark Lindblad and Henry Lindeman and Alex Meyer and Parth Parmar and Tanvi Ranade and Mehul A. Shah and Benjamin Sowell and Dan Tecuci and Vinayak Thapliyal and Matt Welsh},
      year={2024},
      eprint={2409.00847},
      archivePrefix={arXiv},
      primaryClass={cs.DB},
      url={https://arxiv.org/abs/2409.00847}, 
}

@misc{yu2018syntaxsqlnetsyntaxtreenetworks,
      title={{SyntaxSQLNet}: Syntax Tree Networks for Complex and Cross-Domain Text-to-{SQL} Task}, 
      author={Tao Yu and Michihiro Yasunaga and Kai Yang and Rui Zhang and Dongxu Wang and Zifan Li and Dragomir Radev},
      year={2018},
      eprint={1810.05237},
      archivePrefix={arXiv},
      primaryClass={cs.CL},
      url={https://arxiv.org/abs/1810.05237}, 
}

@misc{yu2019spiderlargescalehumanlabeleddataset,
      title={Spider: A Large-Scale Human-Labeled Dataset for Complex and Cross-Domain Semantic Parsing and Text-to-{SQL} Task}, 
      author={Tao Yu and Rui Zhang and Kai Yang and Michihiro Yasunaga and Dongxu Wang and Zifan Li and James Ma and Irene Li and Qingning Yao and Shanelle Roman and Zilin Zhang and Dragomir Radev},
      year={2019},
      eprint={1809.08887},
      archivePrefix={arXiv},
      primaryClass={cs.CL},
      url={https://arxiv.org/abs/1809.08887}, 
}

@misc{guo2019complextexttosqlcrossdomaindatabase,
      title={Towards Complex Text-to-{SQL} in Cross-Domain Database with Intermediate Representation}, 
      author={Jiaqi Guo and Zecheng Zhan and Yan Gao and Yan Xiao and Jian-Guang Lou and Ting Liu and Dongmei Zhang},
      year={2019},
      eprint={1905.08205},
      archivePrefix={arXiv},
      primaryClass={cs.CL},
      url={https://arxiv.org/abs/1905.08205}, 
}

@misc{scholak2021picardparsingincrementallyconstrained,
      title={{PICARD}: Parsing Incrementally for Constrained Auto-Regressive Decoding from Language Models}, 
      author={Torsten Scholak and Nathan Schucher and Dzmitry Bahdanau},
      year={2021},
      eprint={2109.05093},
      archivePrefix={arXiv},
      primaryClass={cs.CL},
      url={https://arxiv.org/abs/2109.05093}, 
}

@misc{gao2024retrievalaugmentedgenerationlargelanguage,
      title={Retrieval-Augmented Generation for Large Language Models: A Survey}, 
      author={Yunfan Gao and Yun Xiong and Xinyu Gao and Kangxiang Jia and Jinliu Pan and Yuxi Bi and Yi Dai and Jiawei Sun and Meng Wang and Haofen Wang},
      year={2024},
      eprint={2312.10997},
      archivePrefix={arXiv},
      primaryClass={cs.CL},
      url={https://arxiv.org/abs/2312.10997}, 
}

@misc{chen2023largelanguagemodelsfew1shot,
      title={Large Language Models are few(1)-shot Table Reasoners}, 
      author={Wenhu Chen},
      year={2023},
      eprint={2210.06710},
      archivePrefix={arXiv},
      primaryClass={cs.CL},
      url={https://arxiv.org/abs/2210.06710}, 
}

@misc{fang2024largelanguagemodelsllmstabular,
      title={Large Language Models ({LLMs}) on Tabular Data: Prediction, Generation, and Understanding -- A Survey}, 
      author={Xi Fang and Weijie Xu and Fiona Anting Tan and Jiani Zhang and Ziqing Hu and Yanjun Qi and Scott Nickleach and Diego Socolinsky and Srinivasan Sengamedu and Christos Faloutsos},
      year={2024},
      eprint={2402.17944},
      archivePrefix={arXiv},
      primaryClass={cs.CL},
      url={https://arxiv.org/abs/2402.17944}, 
}

@misc{li2023tablegpttabletunedgptdiverse,
      title={{Table-GPT}: Table-tuned {GPT} for Diverse Table Tasks}, 
      author={Peng Li and Yeye He and Dror Yashar and Weiwei Cui and Song Ge and Haidong Zhang and Danielle Rifinski Fainman and Dongmei Zhang and Surajit Chaudhuri},
      year={2023},
      eprint={2310.09263},
      archivePrefix={arXiv},
      primaryClass={cs.CL},
      url={https://arxiv.org/abs/2310.09263}, 
}

@misc{liu2025loglmtaskbasedinstructionbasedautomated,
      title={{LogLM}: From Task-based to Instruction-based Automated Log Analysis}, 
      author={Yilun Liu and Yuhe Ji and Shimin Tao and Minggui He and Weibin Meng and Shenglin Zhang and Yongqian Sun and Yuming Xie and Boxing Chen and Hao Yang},
      year={2025},
      eprint={2410.09352},
      archivePrefix={arXiv},
      primaryClass={cs.SE},
      url={https://arxiv.org/abs/2410.09352}, 
}

@misc{cui2024logevalcomprehensivebenchmarksuite,
      title={{LogEval}: A Comprehensive Benchmark Suite for Large Language Models In Log Analysis}, 
      author={Tianyu Cui and Shiyu Ma and Ziang Chen and Tong Xiao and Shimin Tao and Yilun Liu and Shenglin Zhang and Duoming Lin and Changchang Liu and Yuzhe Cai and Weibin Meng and Yongqian Sun and Dan Pei},
      year={2024},
      eprint={2407.01896},
      archivePrefix={arXiv},
      primaryClass={cs.CL},
      url={https://arxiv.org/abs/2407.01896}, 
}

@misc{vaarandi2025usinglargelanguagemodels,
      title={Using Large Language Models for Template Detection from Security Event Logs}, 
      author={Risto Vaarandi and Hayretdin Bahsi},
      year={2025},
      eprint={2409.05045},
      archivePrefix={arXiv},
      primaryClass={cs.CR},
      url={https://arxiv.org/abs/2409.05045}, 
}

@article{gemini,
  author = {Gemini Team and Google},
  title = {Gemini: A Family of Highly Capable Multimodal Models},
  journal = {arXiv preprint arXiv:2312.11805},
  year = {2023},
  url = {https://arxiv.org/abs/2312.11805}
}

@misc{wang2023selfconsistencyimproveschainthought,
      title={Self-Consistency Improves Chain of Thought Reasoning in Language Models}, 
      author={Xuezhi Wang and Jason Wei and Dale Schuurmans and Quoc Le and Ed Chi and Sharan Narang and Aakanksha Chowdhery and Denny Zhou},
      year={2023},
      eprint={2203.11171},
      archivePrefix={arXiv},
      primaryClass={cs.CL},
      url={https://arxiv.org/abs/2203.11171}, 
}

@misc{googleparsers,
  author       = {{Google Cloud}},
  title        = {{Supported Default Parsers | Chronicle Security Operations}},
  year         = {2025},
  url          = {https://cloud.google.com/chronicle/docs/ingestion/parser-list/supported-default-parsers},
  note         = {Accessed: 2025-08-05}
}

@misc{splunkparsers,
  author       = {{Splunk Inc.}},
  title        = {{Add‑ons and CIM | Splunk Supported Add‑ons Documentation}},
  year         = {2025},
  note         = {Accessed: 2025‑08‑05},
  url          = {https://docs.splunk.com/Documentation/AddOns/released/Overview/Add‑onsandCIM}
}

@misc{niccs_nice_framework,
  author       = {{Cybersecurity and Infrastructure Security Agency (CISA)}},
  title        = {{NICE Workforce Framework for Cybersecurity (NICE Framework)}},
  year         = {2025},
  note         = {Last published: May 29, 2025; Accessed: 2025‑08‑05},
  url          = {https://niccs.cisa.gov/tools/nice-framework}
}

@online{udm,
  title   = {{UDM field list}},
  author  = {{Google Cloud}},
  date    = {2025-08-18},
  url     = {https://cloud.google.com/chronicle/docs/reference/udm-field-list},
  urldate = {2025-08-20}
}

@misc{RedHatBugzilla,
  title        = {{Red Hat Bugzilla}},
  year         = {2025},
  organization = {Red Hat, Inc.},
  url          = {https://bugzilla.redhat.com/}
}

@techreport{menlo2012,
  title        = {The {Menlo} Report: Ethical Principles Guiding Information and Communication Technology Research},
  author       = {{U.S. Department of Homeland Security}},
  year         = {2012},
  month        = {August},
  institution  = {Department of Homeland Security},
  url          = {https://www.dhs.gov/sites/default/files/publications/CSD-MenloPrinciplesCORE-20120803_1.pdf}
}

@misc{openai-o4mini-2025,
  title        = {{Introducing OpenAI o3 and o4-mini}},
  author       = {OpenAI},
  year         = {2025},
  url          = {https://openai.com/index/introducing-o3-and-o4-mini/},
  note         = {Accessed: 2025-09-30}
}

@misc{schick2023toolformerlanguagemodelsteach,
      title={Toolformer: Language Models Can Teach Themselves to Use Tools}, 
      author={Timo Schick and Jane Dwivedi-Yu and Roberto Dessì and Roberta Raileanu and Maria Lomeli and Luke Zettlemoyer and Nicola Cancedda and Thomas Scialom},
      year={2023},
      eprint={2302.04761},
      archivePrefix={arXiv},
      primaryClass={cs.CL},
      url={https://arxiv.org/abs/2302.04761}, 
}

@inproceedings{dbscan,
author = {Ester, Martin and Kriegel, Hans-Peter and Sander, J\"{o}rg and Xu, Xiaowei},
title = {A density-based algorithm for discovering clusters in large spatial databases with noise},
year = {1996},
booktitle = {Proceedings of the Second International Conference on Knowledge Discovery and Data Mining},
}


\clearpage
\appendices

\section*{Ethical Considerations}

\smallskip\noindent\textit{Stakeholders.}
Our work potentially impacts multiple groups: (i) security analysts and organizations that rely on log data for detection and forensics, (ii) end users whose activities are indirectly reflected in log data, (iii) service providers that generate and process logs, and (iv) the research community developing AI-driven security analytics.

\smallskip\noindent\textit{Impacts and Principles.}
Following the Menlo Report~\cite{menlo2012}, we considered:

\begin{itemize}[leftmargin=*]
  \item \textbf{Beneficence.} Our goal is to improve security monitoring efficiency by reducing manual parser creation.
  \item \textbf{Respect for Persons.} We only used public datasets (e.g., Red Hat Bugzilla logs) or enterprise logs under strict safeguards.
  \item \textbf{Justice.} Results are shared via open benchmarks and code to ensure broad benefit.
  \item \textbf{Respect for Law and Public Interest.} We adhered to data policies and avoided any live system interaction.
\end{itemize}

\smallskip\noindent\textit{Potential Harms and Mitigations.}
\begin{itemize}[leftmargin=*]
  \item \textbf{Privacy.} We processed enterprise data only on authorized systems and share only aggregate, vetted results.
  \item \textbf{Operational.} Experiments were run offline to avoid production impact.
  \item \textbf{Workforce.} Automation reduces repetitive workload without displacing analysts.
\end{itemize}

We judged the benefits—improved defenses, reduced analyst burden, and reproducibility—to outweigh residual risks.

\section{Additional results}
\label{app:add_results}

We provide here detailed metrics for (1) the template generation metrics for each file in the LogHub2.0 dataset (\cref{ssec:LogHub2.0}), and (2) precision and recall for each file in the SecurityLogs dataset for our ablation studies (\cref{ssec:bugtracker}).

\begin{table*}[b]
\centering
\caption{Comparison of \sys and prior template generation on LogHub2.0 data}
\setlength{\tabcolsep}{5pt} 
\begin{tabular}{|l|rr|rr|rr|rr||l|rr|rr|rr|rr|}
\hline
\multirow{2}{*}{\textbf{Dataset}} & \multicolumn{2}{c|}{\textbf{\sys}} & \multicolumn{2}{c|}{\textbf{LILAC}} & \multicolumn{2}{c|}{\textbf{Drain3}} & \multicolumn{2}{c||}{\textbf{Brain}} & 
\multirow{2}{*}{\textbf{Dataset}} & \multicolumn{2}{c|}{\textbf{\sys}} & \multicolumn{2}{c|}{\textbf{LILAC}} & \multicolumn{2}{c|}{\textbf{Drain3}} & \multicolumn{2}{c|}{\textbf{Brain}} \\
\cline{2-9}\cline{11-18}
& \textbf{PGS} & \textbf{TS} & \textbf{PGS} & \textbf{TS} & \textbf{PGS} & \textbf{TS} & \textbf{PGS} & \textbf{TS} & & \textbf{PGS} & \textbf{TS} & \textbf{PGS} & \textbf{TS} & \textbf{PGS} & \textbf{TS} & \textbf{PGS} & \textbf{TS} \\
\hline
\hline
Apache & {\bf 1.00} & {\bf 1.00} & {\bf 1.00} & {\bf 1.00} & 0.99 & 0.89 & 0.99 & 0.88 & HPC & {\bf 1.00} & {\bf 1.00} & {\bf 1.00} & {\bf 1.00} & 0.97 & 0.96 & 0.78 & 0.75\\
\hline
BGL & {\bf 0.99} & {\bf 0.99} & {\bf 0.99} & {\bf 0.98}  & 0.95 & 0.96 & 0.88 & 0.79 & Linux & 0.87 & 0.92 & {\bf 0.91} & {\bf 0.94} & 0.87 & 0.80 & 0.81 & 0.71 \\
\hline
Hadoop & 0.98 & {\bf 0.94} & {\bf 0.99} & 0.92 & 0.88 & 0.71 & 0.35 & 0.29 & Mac & {\bf 0.96} & {\bf 0.94} & 0.91 & 0.92 & 0.83 & 0.82 & 0.95 & 0.89 \\
\hline
HDFS & {\bf 1.00} & {\bf 0.98} & {\bf 1.00} & 0.87 & 0.97 & 0.93 & 0.97 & 0.84 & OpenSSH & 0.97 & {\bf 0.97} & 0.77 & 0.76 & 0.96 & 0.92 & {\bf 0.99} & 0.92 \\
\hline
HealthApp & {\bf 1.00} & 0.91 & {\bf 1.00} & 0.89 & 0.97 & {\bf 0.92} & 0.54 & 0.44 & OpenStack & 0.88 & 0.88 & 0.83 & 0.81 & 0.66 & 0.67 & {\bf 1.00} & {\bf 0.97} \\
\hline
Proxifier & {\bf 1.00} & {\bf 0.50} & {\bf 1.00} & 0.4 & 0.52 & 0.31 & 0.99 & 0.35 & Spark & 0.97 & 0.95 & {\bf 1.00} & {\bf 0.97} & 0.97 & 0.79 & 0.97 & 0.91\\
\hline
Thunderbird & 0.97 & {\bf 0.86} & {\bf 0.99} & {\bf 0.86} & 0.87 & 0.68 & 0.79 & 0.69 & Zookeeper & {\bf 1.00} & {\bf 0.98} & 0.86 & 0.83 & 0.99 & {\bf 0.98} & 0.99 & 0.95\\
\hline
\end{tabular}
\label{fig:LogHub2.0-full}
\end{table*}

\begin{table*}[b]
\centering
\small
\caption{\sys query precision and recall metrics across Gemini models on SecurityLogs}
\setlength{\tabcolsep}{2.5pt} 
\begin{tabularx}{\textwidth}{|l|Y Y|Y Y|Y Y|Y Y|Y Y|Y Y|Y Y|Y Y|Y Y|Y Y|}
\hline
\multirow{3}{*}{\textbf{Dataset}}
& \multicolumn{4}{c|}{\textbf{Gemini 2.5 Flash~\cite{gemini}}}
& \multicolumn{4}{c|}{\textbf{Gemini 2.5 Pro~\cite{gemini}}}
& \multicolumn{4}{c|}{\textbf{\makecell{Gemini 2.5 Pro \\ No Validation}}}
& \multicolumn{4}{c|}{\textbf{\makecell{Gemini 2.5 Pro \\ No Fewshot}}}
& \multicolumn{4}{c|}{\textbf{o4-mini~\cite{openai-o4mini-2025}}} \\
\cline{2-21}
& \multicolumn{2}{c|}{\textbf{OCSF}} & \multicolumn{2}{c|}{\textbf{Custom}}
& \multicolumn{2}{c|}{\textbf{OCSF}} & \multicolumn{2}{c|}{\textbf{Custom}}
& \multicolumn{2}{c|}{\textbf{OCSF}} & \multicolumn{2}{c|}{\textbf{Custom}}
& \multicolumn{2}{c|}{\textbf{OCSF}} & \multicolumn{2}{c|}{\textbf{Custom}}
& \multicolumn{2}{c|}{\textbf{OCSF}} & \multicolumn{2}{c|}{\textbf{Custom}} \\
\cline{2-21}
& \textbf{\small P} & \textbf{\small R}
& \textbf{\small P} & \textbf{\small R}
& \textbf{\small P} & \textbf{\small R}
& \textbf{\small P} & \textbf{\small R}
& \textbf{\small P} & \textbf{\small R}
& \textbf{\small P} & \textbf{\small R}
& \textbf{\small P} & \textbf{\small R}
& \textbf{\small P} & \textbf{\small R}
& \textbf{\small P} & \textbf{\small R}
& \textbf{\small P} & \textbf{\small R} \\
\hline
\hline

SSHD
& 0.70 & 0.70 & \textbf{1.00} & \textbf{1.00}
& 0.90 & 0.90 & \textbf{1.00} & \textbf{1.00}
& 0.80 & 0.80 & \textbf{1.00} & \textbf{1.00}
& 0.40 & 0.40 & 0.90 & 0.81
& \textbf{1.00} & \textbf{1.00} & \textbf{1.00} & \textbf{1.00} \\
\hline

Cron
& \textbf{1.00} & \textbf{1.00} & \textbf{1.00} & \textbf{1.00}
& \textbf{1.00} & \textbf{1.00} & \textbf{1.00} & \textbf{1.00}
& \textbf{1.00} & \textbf{1.00} & \textbf{1.00} & \textbf{1.00}
& 0.40 & 0.40 & 0.60 & 0.60
& 0.80 & 0.80 & \textbf{1.00} & \textbf{1.00} \\
\hline

DHCP
& 0.80 & 0.77 & 0.90 & 0.87
& 0.80 & 0.77 & \textbf{1.00} & \textbf{0.97}
& 0.54 & 0.57 & 0.94 & \textbf{0.97}
& 0.41 & 0.47 & 0.81 & 0.82
& 0.64 & 0.62 & 0.90 & 0.83 \\
\hline

Audit
& 0.60 & 0.60 & \textbf{1.00} & 0.96
& 0.70 & 0.69 & \textbf{1.00} & \textbf{0.99}
& 0.70 & 0.68 & \textbf{1.00} & 0.95
& 0.20 & 0.19 & 0.70 & 0.63
& 0.80 & 0.80 & \textbf{1.00} & 0.98 \\
\hline

Puppet
& 0.70 & 0.71 & 0.70 & 0.71
& 0.90 & 0.88 & \textbf{1.00} & \textbf{0.98}
& 0.90 & 0.79 & \textbf{1.00} & 0.89
& 0.69 & 0.69 & 0.69 & 0.69
& 0.89 & 0.85 & \textbf{1.00} & 0.95 \\
\hline
\hline

\rowcolor[gray]{0.9}
Average
& 0.76 & 0.76 & 0.92 & 0.91
& 0.86 & 0.85 & \textbf{1.00} & \textbf{0.99}
& 0.79 & 0.77 & 0.99 & 0.96
& 0.42 & 0.43 & 0.74 & 0.71
& 0.83 & 0.81 & 0.98 & 0.95 \\
\hline

\end{tabularx}
\label{tab:metrics-query-ablation-full}
\end{table*}

\section{Evaluation queries}
\label{app:queries}

\begin{table*}[b]
\caption{DHCP Log Queries}
\centering
\begin{tabularx}{\textwidth}{|p{0.25\textwidth}|p{0.15\textwidth}|X|}
\hline
\textbf{Assigned addresses for a given hostname} & Description & Identifies IP addresses assigned to a specific host. \\
\cline{2-3}
 & Query & \texttt{assigned\_ip exists AND log\_host is laphroaig} \\
\cline{2-3}
 & Naive substring & \texttt{fgrep "bound to" | fgrep "laphroaig"} \\
\hline
\textbf{List all server IPs} & Description & Enumerates all DHCP server IP addresses that are not broadcast addresses. \\
\cline{2-3}
 & Query & \texttt{server\_ip != 255.255.255.255} \\
\cline{2-3}
 & Naive substring & \texttt{fgrep "port" | fgrep -v "255.255.255.255"} \\
\cline{2-3}
 & Note & The word port seems to always be present next to server IPs \\
\hline
\textbf{Track all log messages for a given transaction ID} & Description & Follows the complete lifecycle of a specific DHCP transaction. \\
\cline{2-3}
 & Query & \texttt{transaction\_id is 0x6520bf0e} \\
\cline{2-3}
 & Naive substring & \texttt{fgrep "xid=0x6520bf0e"} \\
\hline
\textbf{List the MAC addresses used by a specific host} & Description & Retrieves entries containing MAC address information for a particular hostname. \\
\cline{2-3}
 & Query & \texttt{mac\_address exists and log\_host is laphroaig} \\
\cline{2-3}
 & Naive substring & \texttt{fgrep "laphroaig" | fgrep "Listening"} \\
\hline
\textbf{Entries with high renewal times} & Description & Identifies DHCP leases with renewal times exceeding 1 day (86400 seconds). \\
\cline{2-3}
 & Query & \texttt{renewal\_time > 86400} \\
\cline{2-3}
 & Naive substring & \texttt{fgrep "renewal"} \\
\cline{2-3}
 & Note & We cannot compare numbers so we can only search for lines that include renewal times. \\
\hline
\textbf{Servers on non-standard ports} & Description & Lists DHCP servers operating on ports other than the standard port 67. \\
\cline{2-3}
 & Query & \texttt{server\_port is not 67} \\
\cline{2-3}
 & Naive substring & \texttt{fgrep "port" | fgrep -v "67"} \\
\hline
\textbf{Specific client version usage} & Description & Identifies log entries associated with a specific DHCP client version (3.0.1). \\
\cline{2-3}
 & Query & \texttt{client\_version is 3.0.1} \\
\cline{2-3}
 & Naive substring & \texttt{fgrep "3.0.1"} \\
\hline
\textbf{DHCPDISCOVER messages} & Description & Lists clients that have sent DHCPDISCOVER messages. \\
\cline{2-3}
 & Query & \texttt{DHCP\_message\_type is DHCPDISCOVER} \\
\cline{2-3}
 & Naive substring & \texttt{fgrep "DHCPDISCOVER"} \\
\hline
\textbf{XMT Renew messages} & Description & Lists clients that have issued renewal requests. \\
\cline{2-3}
 & Query & \texttt{DHCP\_message\_type is Renew} \\
\cline{2-3}
 & Naive substring & \texttt{fgrep "Renew"} \\
\hline
\textbf{Bad IP checksums} & Description & Identifies packets with incorrect IP checksums. \\
\cline{2-3}
 & Query & \texttt{bad IP checksums} \\
\cline{2-3}
 & Naive substring & \texttt{fgrep "bad IP checksums"} \\
\hline
\end{tabularx}
\end{table*}

\clearpage

\begin{table*}[H]
\caption{SSHD Log Queries}
\centering
\begin{tabularx}{\textwidth}{|p{0.25\textwidth}|p{0.15\textwidth}|X|}
\hline
\textbf{Password authentication for root} & Description & Identifies instances where the root account attempted to authenticate using a password. \\
\cline{2-3}
 & Query & \texttt{authentication\_method is password and user\_name is root} \\
\cline{2-3}
 & Naive substring & \texttt{fgrep "password" | fgrep "root"} \\
\hline
\textbf{Unusual server ports} & Description & Detects SSH servers operating on non-standard ports (not port 22). \\
\cline{2-3}
 & Query & \texttt{bind\_port is not 22} \\
\cline{2-3}
 & Naive substring & \texttt{fgrep "port" | fgrep -v "22"} \\
\hline
\textbf{Usage of specific key} & Description & Tracks usage of a particular SSH key based on its fingerprint. \\
\cline{2-3}
 & Query & \texttt{key\_hash is SHA256:iJC3O+heWZCsp5$\cdots$+VwGmmcFhEnc} \\
\cline{2-3}
 & Naive substring & \texttt{fgrep "SHA256:iJC3O+heWZCsp5$\cdots$VwGmmcFhEnc"} \\
\hline
\textbf{Root user SSH keys} & Description & Retrieves all SSH key fingerprints associated with root user logins. \\
\cline{2-3}
 & Query & \texttt{key\_hash exists and user\_name is root} \\
\cline{2-3}
 & Naive substring & \texttt{fgrep "root" | fgrep "publickey"} \\
\hline
\textbf{Activity from specific IP on specific date} & Description & Monitors all SSH activity from IP 61.143.236.193 on September 25. \\
\cline{2-3}
 & Query & \texttt{remote\_ip is 61.143.236.193 and log\_timestamp > sept. 25 00:00:00 and log\_timestamp < sept. 26 00:00:00} \\
\cline{2-3}
 & Naive substring & \texttt{fgrep "61.143.236.193" | fgrep "sept. 25"} \\
\hline
\textbf{Non-SSH terminals for root user} & Description & Identifies root logins through terminal types other than SSH. \\
\cline{2-3}
 & Query & \texttt{terminal\_type is not ssh and user\_name is root} \\
\cline{2-3}
 & Naive substring & \texttt{fgrep "root" | fgrep "tty" | fgrep -v "tty=ssh"} \\
\hline
\textbf{Root sessions initiated by non-root accounts} & Description & Detects when standard users escalate to root privileges. \\
\cline{2-3}
 & Query & \texttt{initiating\_user\_name is not root and user\_name is root} \\
\cline{2-3}
 & Naive substring & \texttt{fgrep "user root" | fgrep -v "root("} \\
\cline{2-3}
 & Note & This matches the format of the main template that contains this information \\
\hline
\textbf{``None'' authentication attempts} & Description & Identifies login attempts using the "none" authentication method. \\
\cline{2-3}
 & Query & \texttt{authentication\_method is none} \\
\cline{2-3}
 & Naive substring & \texttt{fgrep "none"} \\
\hline
\textbf{Activity for specific system and process} & Description & Tracks SSH activity related to a specific process ID on a particular host. \\
\cline{2-3}
 & Query & \texttt{process\_id is 4317 and log\_host is LIPC003.intranet.local} \\
\cline{2-3}
 & Naive substring & \texttt{fgrep "4317" | fgrep "LIPC003.intranet.local"} \\
\hline
\textbf{Host key mentions} & Description & Finds log entries that reference host key files or paths. \\
\cline{2-3}
 & Query & \texttt{host\_key\_path exists} \\
\cline{2-3}
 & Naive substring & \texttt{fgrep "host key"} \\
\hline
\end{tabularx}
\end{table*}

\clearpage

\begin{minipage}{\textwidth}
\begin{table}[H]
\caption{Audit Log Queries}
\centering
\begin{tabularx}{\textwidth}{|p{0.25\textwidth}|p{0.15\textwidth}|X|}
\hline
\textbf{Sudo/su usage by non-root users} & Description & Identifies when standard users attempt to use sudo or su commands. \\
\cline{2-3}
 & Query & \texttt{(executable\_path contains /sudo or executable\_path contains /su) AND user\_id is not 0} \\
\cline{2-3}
 & Naive substring & \texttt{fgrep "/su" | fgrep -v "uid=0"} \\
\hline
\textbf{Denied write operations for rsync} & Description & Detects when rsync processes are denied write permissions. \\
\cline{2-3}
 & Query & \texttt{avc\_operation contains write and process\_name contains rsync} \\
\cline{2-3}
 & Naive substring & \texttt{fgrep "write" | fgrep "rsync"} \\
\hline
\textbf{Specific Target SELinux context} & Description & Finds log entries with a specific target SELinux context. \\
\cline{2-3}
 & Query & \texttt{\footnotesize target\_context is system\_u:system\_r:udev\_t:s0-s0:c0.c1023} \\
\cline{2-3}
 & Naive substring & \texttt{fgrep "system\_u:system\_r:udev\_t:s0-s0:c0.c1023"} \\
\hline
\textbf{Devices that had denied calls to mount} & Description & Identifies log entries related to mounting storage devices. \\
\cline{2-3}
 & Query & \texttt{device\_name exists and process\_name is "mount"} \\
\cline{2-3}
 & Naive substring & \texttt{fgrep "mount" | fgrep 'dev='} \\
\hline
\textbf{Root user logins} & Description & Captures all direct login events for the root user. \\
\cline{2-3}
 & Query & \texttt{audit\_type is LOGIN and user\_id is 0} \\
\cline{2-3}
 & Naive substring & \texttt{fgrep "LOGIN" | fgrep "uid=0"} \\
\hline
\textbf{Audit rule removal} & Description & Detects when audit rules are removed from the system. \\
\cline{2-3}
 & Query & \texttt{operation contains "remove rule"} \\
\cline{2-3}
 & Naive substring & \texttt{fgrep "remove rule"} \\
\hline
\textbf{SELinux permissive mode setting} & Description & Identifies when SELinux is set to permissive mode rather than enforcing. \\
\cline{2-3}
 & Query & \texttt{selinux\_permissive is 1} \\
\cline{2-3}
 & Naive substring & \texttt{fgrep "permissive=1"} \\
\hline
\textbf{Root directory as working directory} & Description & Finds processes operating with root (/) as their current working directory. \\
\cline{2-3}
 & Query & \texttt{current\_working\_directory is '/' or current\_working\_directory is '"/"'} \\
\cline{2-3}
 & Naive substring & \texttt{fgrep "cwd=/ " and fgrep 'cwd="/"'} \\
\hline
\textbf{Non-binary audit enabled flags} & Description & Detects when the audit enabled flag is set to a value other than 0 or 1. \\
\cline{2-3}
 & Query & \texttt{audit\_enabled is not 0 and audit\_enabled is not 1} \\
\cline{2-3}
 & Naive substring & \texttt{fgrep "audit\_enabled=" | fgrep -v "audit\_enabled=1" | fgrep -v "audit\_enabled=0"} \\
\hline
\textbf{Remote SSH connections to specific host on specific date} & Description & Tracks remote hosts that established SSH connections to a particular server on a specific date. \\
\cline{2-3}
 & Query & \texttt{audit\_datetime >= Aug 3 and audit\_datetime < Aug 4 and terminal contains ssh and remote\_hostname exists and audit\_host is perfc-380g8-01} \\
\cline{2-3}
 & Naive substring & \texttt{fgrep "perfc-380g8-01" | fgrep "Aug  3" | fgrep "terminal=ssh" | fgrep "hostname"} \\
\hline
\end{tabularx}
\end{table}
\end{minipage}

\clearpage
\begin{minipage}{\textwidth}
\begin{table}[H]
\caption{Cron Log Queries}
\centering
\begin{tabularx}{\textwidth}{|p{0.25\textwidth}|p{0.15\textwidth}|X|}
\hline
\textbf{List all executed jobs on a host at a specific date} & Description & Identifies entries with executable paths on a particular host within a specific date range. \\
\cline{2-3}
 & Query & \texttt{executable\_path exists and log\_timestamp >= 2017-07-14 and log\_timestamp < 2017-07-15 and log\_hostname is httpboot} \\
\cline{2-3}
 & Naive substring & \texttt{fgrep "CMD" | fgrep "2017-07-14" | fgrep "httpboot"} \\
\hline
\textbf{Entries with scaling factor} & Description & Locates log entries that contain scaling factor information. \\
\cline{2-3}
 & Query & \texttt{scaling\_factor exists} \\
\cline{2-3}
 & Naive substring & \texttt{fgrep "factor"} \\
\hline
\textbf{Specific process ID} & Description & Finds entries related to a specific process ID. \\
\cline{2-3}
 & Query & \texttt{process\_id is 24225} \\
\cline{2-3}
 & Naive substring & \texttt{fgrep "24225"} \\
\hline
\textbf{CRON session openings for root} & Description & Lists all session openings for the root user before a specific date. \\
\cline{2-3}
 & Query & \texttt{opened and username is root and log\_timestamp < 2017-07-15} \\
\cline{2-3}
 & Naive substring & \texttt{fgrep "opened" | fgrep "root" | fgrep "2017-07-14"} \\
\cline{2-3}
 & Note & We cannot compare dates so we look for the day before \\
\hline
\textbf{CRON session closings} & Description & Lists all session closings before a specific date. \\
\cline{2-3}
 & Query & \texttt{closed and log\_timestamp < 2017-07-15} \\
\cline{2-3}
 & Naive substring & \texttt{fgrep "closed" | fgrep "2017-07-14"} \\
\hline
\end{tabularx}
\end{table}
\end{minipage}

\clearpage
\begin{minipage}{\textwidth}
\begin{table}[H]
\caption{Puppet Log Queries}
\centering
\begin{tabularx}{\textwidth}{|p{0.25\textwidth}|p{0.15\textwidth}|X|}
\hline
\textbf{Resource-specific failures for a host} & Description & Retrieves failure reports about a specific Puppet resource on a particular host. \\
\cline{2-3}
 & Query & \texttt{puppet\_resource is Service[galera] AND has failures AND log\_hostname is controller1} \\
\cline{2-3}
 & Naive substring & \texttt{fgrep "Service[galera]" | fgrep "has failures" | fgrep "controller1"} \\
\hline
\textbf{Revoked certificates} & Description & Identifies logs reporting revoked certificates. \\
\cline{2-3}
 & Query & \texttt{certificate\_common\_name exists AND revoked} \\
\cline{2-3}
 & Naive substring & \texttt{fgrep "revoked"} \\
\hline
\textbf{Specific error code on similar hosts} & Description & Finds hosts with similar naming patterns experiencing a specific error code. \\
\cline{2-3}
 & Query & \texttt{error\_code is 14 and log\_hostname contains maca} \\
\cline{2-3}
 & Naive substring & \texttt{fgrep "14" | fgrep "maca"} \\
\hline
\textbf{Host associated with specific request ID} & Description & Identifies the host linked to a particular Puppet request ID. \\
\cline{2-3}
 & Query & \texttt{\footnotesize request\_id is req-9ac8edb7-f81f-44a7-9f34-9a375e7df573} \\
\cline{2-3}
 & Naive substring & \texttt{fgrep "req-9ac8edb7-f81f-44a7-9f34-9a375e7df573"} \\
\hline
\textbf{Interval value changes} & Description & Tracks changes to interval values in Puppet configurations. \\
\cline{2-3}
 & Query & \texttt{attribute\_name is interval and new\_value exists} \\
\cline{2-3}
 & Naive substring & \texttt{fgrep "interval"} \\
\hline
\textbf{Specific SQL password hash detection} & Description & Checks if any SQL-related resources contain a specific password hash value. \\
\cline{2-3}
 & Query & \texttt{attribute\_name is password\_hash and new\_value contains D602AB02F4227D3EBF5FE6EA0323BD6D586A7454 and reporting\_resource contains sql} \\
\cline{2-3}
 & Naive substring & \texttt{fgrep "D602AB02F4227D3EBF5FE6EA0323BD6D586A7454" | fgrep "sql"} \\
\hline
\textbf{Extended Puppet run durations} & Description & Identifies Puppet runs that took longer than 1 hour (3600 seconds). \\
\cline{2-3}
 & Query & \texttt{run\_time > 3600} \\
\cline{2-3}
 & Naive substring & \texttt{fgrep "catalog run"} \\
\cline{2-3}
 & Note & We cannot compare numbers without parsing \\
\hline
\textbf{Non-localhost server connections} & Description & Lists connections to non-localhost servers by a Puppet agent on a specific date. \\
\cline{2-3}
 & Query & \texttt{server\_ip is not 127.0.0.1 and log\_hostname is puma03 and log\_timestamp >= Jan 8 and log\_timestamp < Jan 9} \\
\cline{2-3}
 & Naive substring & \texttt{fgrep "127.0.0.1" | fgrep "puma03" | fgrep "Jan  8"} \\
\hline
\textbf{Firewall persistence failures} & Description & Identifies cases where firewall rules cannot be persisted. \\
\cline{2-3}
 & Query & \texttt{Unable to persist firewall rules} \\
\cline{2-3}
 & Naive substring & \texttt{fgrep "Unable to persist firewall rules"} \\
\hline
\textbf{HTTP URL targets} & Description & Lists log entries with HTTP URL targets. \\
\cline{2-3}
 & Query & \texttt{target\_url contains http://} \\
\cline{2-3}
 & Naive substring & \texttt{fgrep "http://"} \\
\hline
\end{tabularx}
\end{table}
\end{minipage}

\end{document}